\documentclass[prd,aps,nofootinbib,floatfix,11pt]{revtex4}
\usepackage{amsmath,graphicx,epsfig,amssymb,dsfont,mathtools,}
\usepackage[usenames]{color}
\usepackage{ulem} 
\usepackage{bigstrut}
\usepackage{slashed}
\usepackage{multirow}
\usepackage{subfigure}

\allowdisplaybreaks

\begin{document}
\title{Semi-inclusive decays of $B$ meson into a dark anti-baryon and baryons}

\author{Yu-Ji Shi$^{1,4}$~\footnote{Email:shiyuji@ecust.edu.cn},
  Ye Xing~$^{2}$~\footnote{Email:xingye\_guang@cumt.edu.cn}
  and Zhi-Peng Xing~$^{3,5}$~\footnote{Email:zpxing@sjtu.edu.cn}}
\affiliation{$^1$ School of Physics, East China University of Science and Technology, Shanghai 200237, China\\
$^2$ School of Materials Science and Physics, China University of Mining and Technology, Xuzhou 221116, China\\
$^{3}$ Tsung-Dao Lee Institute, Shanghai Jiao Tong University, Shanghai 200240, China\\
$^{4}$ Shanghai Key Laboratory of Particle Physics and Cosmology, \\
School of Physics and Astronomy, Shanghai Jiao Tong University, Shanghai 200240, China\\
$^{5}$ Department of Physics and Institute of Theoretical Physics, Nanjing Normal University, Nanjing, Jiangsu 210023, China}

\begin{abstract}
Using the recently developed $B$-Mesogenesis scenario, we studied the semi-inclusive decays of $B$ meson into a dark anti-baryon $\psi$ plus any possible states $X$ containing $u/c$ and $d/s$ quarks with unit baryon number. The two types of effective Lagrangians proposed by the scenario are both considered in the study. The semi-inclusive decay branching fractions of $B\to X \psi$  are calculated by the method of heavy quark expansion,  where the non-perturbative contributions from the matrix elements of dimension-5 operators are included. We obtained the branching fractions as functions of the dark anti-baryon mass. Using the experimental  upper limits of the branching fractions, we presented the constraints of the coupling constants in the  $B$-Mesogenesis scenario.
\end{abstract}
\maketitle

\section{Introduction}
The Standard Model of particle physics and the standard cosmological model are two highly successful frameworks for describing the most microscopic and macroscopic physics respectively. However, these two models are not consistent with each other, which leaves many unanswered questions including the existence of dark matter and the asymmetry of matter and anti-matter. To answer these questions, many mechanisms have been proposed since Sakharov firstly introduced
the conditions necessary for baryogenesis \cite{Sakharov}. The traditional mechanisms generally include high scales and extremely massive particles which makes them difficult to be tested by experiments. Recently, a new $B$-Mesogenesis scenario is proposed by Refs.~\cite{Elor:2018twp,Alonso-Alvarez:2021qfd,Elahi:2021jia}, which can simultaneously explain the relic dark matter abundance and the baryon asymmetry and  in our Universe. The main advantage of this scenario is that it is not only directly testable at hadron colliders and $B$-factories \cite{Alonso-Alvarez:2021qfd,Borsato:2021aum}, but also indirectly testable at Kaon and Hyperon factories \cite{Alonso-Alvarez:2021oaj,Goudzovski:2022vbt}. Nowadays, the search for $B$ meson decays into baryon with missing energy through $B$-Mesogenesis has been independently started by the  Belle-II collaboration \cite{Belle:2021gmc} and the LHCb collaboration \cite{Rodriguez:2021urv}.

In the $B$-Mesogenesis scenario, a new mechanism for Baryogenesis and DM production is proposed. The $b,\bar b$ quarks  are produced by decays of some heavy scalar field $\Phi$ during a late era in the history of the early universe. The produced $b,\bar b$ quarks  hadronize to charged and neutral $B$-mesons. The neutral ones $B^0, {\bar B}^0$ quickly undergo CP violating oscillations, and then decay into a dark sector baryon with baryon number $-1$ as well as visible hadron states with baryon number $+1$.  As a result, the asymmetry of baryon and anti-baryon is produced in the $B$-Mesogenesis without violating the baryon number. The exclusive decay $B\to p \psi$ in the framework of $B$-Mesogenesis was firstly studied by Ref.\cite{Khodjamirian:2022vta} using light-cone sum rules (LCSR). After that, with the use of LCSR, a more complete study of $B$ meson decays into an octet baryon or charmed anti-triplet baryon and $\psi$ was given by Ref.\cite{Elor:2022jxy}. In addition, similar exclusive decays of $B$ meson into a baryon plus missing energy are studied by Ref.\cite{Dib:2022ppx} for probing the lightest neutralino.

Recently, there are no strict theoretical studies on inclusive $B$ meson decays in the $B$-Mesogenesis. Compared with the exclusive decays, inclusive decay branching fractions are more likely to be measured in the experiments. On the other hand, from the theoretical point of view, another advantage of inclusive decays is that the summation over various of hadronic final states eliminates bound-state effects of individual hadrons, which is due to the hypothesis of quark–hadron duality \cite{Poggio:1975af}. In Ref.\cite{Alonso-Alvarez:2021qfd}, using the data of bottom hadron decays with missing energy from the ALEPH experiment \cite{ALEPH:2000vvi,ALEPH:1992zwu,ALEPH:1994bih}, the authors obtained the upper limits on the inclusive decay branching fractions of $B\to X_{u/c,d/s} \psi$, where $X_{u/c,d/s}$ denotes any possible hadron states containing $u/c$ and $d/s$ quarks with unit baryon number. Therefore, compared with the experimental  upper limits, a strict theoretical calculation on the $B\to X \psi$ branching fraction enables us to determine the upper limits on the coupling constants in the $B$-Mesogenesis. Nowadays the heavy quark expansion (HQE)\cite{Manohar:1993qn,Lenz:2014jha,Neubert:1997gu,Falk:1996kh} has been successfully applied for the studies of inclusive decays as well as lifetime calculations of heavy hadron decays \cite{Chay:1990da,Bigi:1992su,Bigi:1993fe,Blok:1993va,Cheng:2018mwu,Kirk:2017juj,Lenz:2013aua,Mannel:2023zei,Piscopo:2023jnu,Mannel:2020fts,Huber:2020vup,Huber:2019iqf,Huber:2018gii,Qin:2021zqx,Yang:2022nps}. In this work, we will use HQE  to calculate the inclusive decay branching fractions of $B\to X_{u/c,d/s} \psi$, where the bound-state effects related to the initial state can be can systematically accounted for by introducing matrix elements of high dimension operators. 

This article is organized as follows: Section II is a brief introduction to the $B$-Mesogenesis scenario proposed by Refs.~\cite{Elor:2018twp,Alonso-Alvarez:2021qfd,Elahi:2021jia}. Section III present a detailed HQE calculation for the $B\to X_{u/c,d/s} \psi$ decays. Section IV gives the numerical results for decay branching fractions and  constraints on the coupling constants in the $B$-Mesogenesis.

\section{$B$-Mesogenesis scenario}
The $B$-Mesogenesis scenario firstly proposed by Refs.~\cite{Elor:2018twp,Alonso-Alvarez:2021qfd,Elahi:2021jia} aims to simultaneously explain the baryon asymmetry and the existence of dark matter in our Universe. This B-Mesogenesis model offers a mechanism where an anti-$b$ quark can decays into $u/c, d/s$ quarks and a dark anti-baryon $\psi$.  Although the baryon number is conserved, $\psi$ is invisible so that only the baryons composed of $u, d/s$ quarks can be detected by the experiments. In Refs.~\cite{Elor:2018twp,Alonso-Alvarez:2021qfd}, such baryon number violating decays are realized by exchanging a charged color triplet scalar $Y^i$. There two types of effective Lagrangians in the $B$-Mesogenesis model with the charge of $Y^i$ being $Q_Y=-1/3$:
\begin{align}
\mathcal{L}_{\rm eff}^{I}=&-y_{ub}\epsilon_{ijk}Y^{*i}{\bar u}_R^j b_R^{c,k}-y_{cb}\epsilon_{ijk}Y^{*i}{\bar c}_R^j b_R^{c,k}-y_{\psi d}Y_i {\bar \psi}d_R^{c,i}-y_{\psi s}Y_i {\bar \psi}s_R^{c,i}+\rm h.c,\nonumber\\
\mathcal{L}_{\rm eff}^{II}=&-y_{ud}\epsilon_{ijk}Y^{*i}{\bar u}_R^j d_R^{c,k}-y_{us}\epsilon_{ijk}Y^{*i}{\bar u}_R^j s_R^{c,k}-y_{cd}\epsilon_{ijk}Y^{*i}{\bar c}_R^j d_R^{c,k}-y_{cs}\epsilon_{ijk}Y^{*i}{\bar c}_R^j s_R^{c,k}\nonumber\\
&-y_{\psi b}Y_i {\bar \psi}b_R^{c,i}+\rm h.c,\label{eq:effLagran1and2}
\end{align}
where all the quark fields are taken as right handed and the superscript $c$ indicates charge conjugate. $Y$ is assumed to be heavy with its mass denoted as $M_Y$. The $y$ s are unknown coupling constants. In the Type-I model the $b$ quark couples with $u,c$ quarks, while in the Type-II model the $b$ quark couples with the dark anti-baryon $\psi$. It should be mentioned that in Ref.~\cite{Alonso-Alvarez:2021qfd} there is a third type of effective Lagrangian with $Q_Y=2/3$, which reads as
\begin{align}
\mathcal{L}_{\rm eff}^{III}=&-y_{bd}\epsilon_{ijk}Y^{*i}{\bar b}_R^j d_R^{c,k}-y_{bs}\epsilon_{ijk}Y^{*i}{\bar b}_R^j s_R^{c,k}-y_{\psi u}Y_i {\bar \psi}u_R^{c,i}-y_{\psi c}Y_i {\bar \psi}c_R^{c,i}+\rm h.c.
\end{align}
In this work, for simplicity we will only consider the case of $Q_Y=-1/3$, which is consistent with the exclusive decay studies in Refs.~\cite{Khodjamirian:2022vta,Elor:2022jxy}.  Integrating out the heavy boson $Y$ in Eq.~(\ref{eq:effLagran1and2}), one arrives at the effective Hamiltonian for the two types of models as:
\begin{align}
&\mathcal{H}_{\rm eff}^{I,uq}=-\frac{y_{ub}y_{\psi q}}{M_Y^2}i\epsilon_{ijk}({\bar \psi} q_R^{c,i})({\bar u}_R^j b_R^{c,k})=-G_{(uq)}^{I}{\bar {\cal O}}_{(uq)}^{I}\psi^c,\nonumber\\
&\mathcal{H}_{\rm eff}^{II,uq}=-\frac{y_{\psi b}y_{u q}}{M_Y^2}i\epsilon_{ijk}({\bar \psi} b_R^{c,i})({\bar u}_R^j q_R^{c,k})=-G_{(uq)}^{II}{\bar {\cal O}}_{(uq)}^{II}\psi^c.\label{eq:effHbuq}
\end{align}
Here for simplicity, $q=s,d$ and $u$ denotes $u$ or $c$ quark simultanously. We have defined three-quark operatora ${\bar {\cal O}}_{(q)}^{I}=-i\epsilon_{ijk}({\bar b}_R^i u_R^{c,j}){\bar q}_R^{k}$ and ${\bar {\cal O}}_{(q)}^{II}=-i\epsilon_{ijk}({\bar q}_R^i u_R^{c,j}){\bar b}_R^{k}$, which transform an anti-$b$ quark into two light quarks $u, q$. In this work, we will calculate the semi-inclusive decay width of $B\to X_{uq}\psi$ induced by $\mathcal{H}_{\rm eff}^{I,uq}$ and $\mathcal{H}_{\rm eff}^{II,uq}$ respectively, with $X_{uq}$ being the summation of any states containing $u,q$ quarks.

\section{$B\to X_{uq}\psi$ decay in heavy quark expansion}
\subsection{Differential decay width of $B\to X_{uq}\psi$}
In the rest frame of $B$ meson, denoting the momentum and energy of the outgoing dark anti-baryon $\psi$ as $q$ and $E$, we can express the differential decay width of $B\to X_{uq}\psi$ as
\begin{align}
\frac{d}{dE}\Gamma(b\to u q \psi)=&\int \frac{d^4 q}{(2\pi)^4}(2\pi)\delta(q^2-m_{\psi}^2)\delta(E-q^0)\nonumber\\
&\times \sum_{X,s_{\psi}}\frac{1}{2 m_B}|\langle X(p_X)\psi(q,s_{\psi})|\mathcal{H}_{\rm eff}^{uq}(0)|B(p_B)\rangle|^2 (2\pi)^4 \delta^4(p_B-q-p_X)\label{eq:diffWidth},
\end{align}
where the spin of $\psi$ and any possible $X_{uq}$ states with momentum $p_X$ are summed. The integration of $E$ is equivalent to averaging over a range of final-state hadronic masses. Since $\psi$ has no strong interaction with quarks, the matrix element in Eq.~(\ref{eq:diffWidth}) can be factoraized as 
\begin{align}
\langle X(p_X)\psi(q,s_{\psi})|\mathcal{H}_{\rm eff}^{uq}(0)|B(p_B)\rangle = -G_{(uq)}\langle X(p_X)|{\bar {\cal O}}_{(uq),a}(0)|B(p_B)\rangle u_{\psi,a}^c(q,s_{\psi}),
\end{align}
with $a$ being a spinor index. For simplicity we have omitted the superscripts $I,II$ here. Now we introduce a rank-two tensor $W$ with two spinor indexes:
\begin{align}
W_{ba}=&\sum_X (2\pi)^3 \delta^4(p_B-q-p_X)\frac{1}{2m_B}\langle B(p_B)|{\bar {\cal O}}^{\dagger}_{(uq),b}(0)|X(p_X)\rangle \langle X(p_X)|{\bar {\cal O}}_{(uq),a}(0)|B(p_B)\rangle,
\end{align}
which can be generally parameterized as
\begin{align}
W=\gamma^0 \left[A_1 \frac{\slashed q}{m_B}+A_2 \frac{\slashed p_B}{m_B}\right]P_L.\label{eq:parameterizeW}
\end{align}
Note that the appearance of $P_L$ on the right hand side is due to the identity ${\bar {\cal O}}_{(uq)}P_R=0$. Now the differential decay width can be expressed in terms of $W$ or $A_{1,2}$ as
\begin{align}
\frac{d}{dE}\Gamma(b\to u q\psi)=&\frac{G_{(q)}^2}{(2\pi)^2}\int d^4 q\  \delta(q^2-m_{\psi}^2)\delta(E-q^0) {\rm Tr}\left[(\slashed q -m_{\psi}) \gamma^0 W\right]\nonumber\\
=&\frac{G_{(s)}^2}{\pi\  m_B}\sqrt{E^2-m_{\psi}^2}\left[A_1(m_{\psi},E)m_{\psi}^2+A_2(m_{\psi},E) m_B E\right].\label{eq:diffWidth2}
\end{align}

It is difficult to calculate the $W$ tensor directly due to the infinite summation on the $X_{uq}$ states. Actually, the $W$ tensor can be extracted from the imaginary part of a correlation function: 
\begin{align}
W_{ba}=-\frac{1}{\pi}{\rm Im}T_{ba}
\end{align}
with
\begin{align}
T_{ba}=&-i\int d^4 x\  e^{-iq\cdot x}\frac{1}{2m_B}\langle B(p_B)|T\left\{{\bar {\cal O}}^{\dagger}_{(uq),b}(x){\bar {\cal O}}_{(uq),a}(0)\right\}|B(p_B)\rangle.\label{eq:Ttensor}
\end{align}
The correlation function defined in Eq.~(\ref{eq:Ttensor}) can be calculated by HQE, where it is expanded according to the power of $1/m_b$. Each term in the expansion is factorized into perturbative part and non-perturbative part. The former one can be calculated perturbatively, while the later one are parameterized by matrix elements of $B$ meson. We will perform an explicit calculation of $T_{ba}$ by HQE in the next section.

\subsection{Heavy quark expansion in the Type-I model}

\begin{figure}
\begin{center}
\includegraphics[width=0.6\columnwidth]{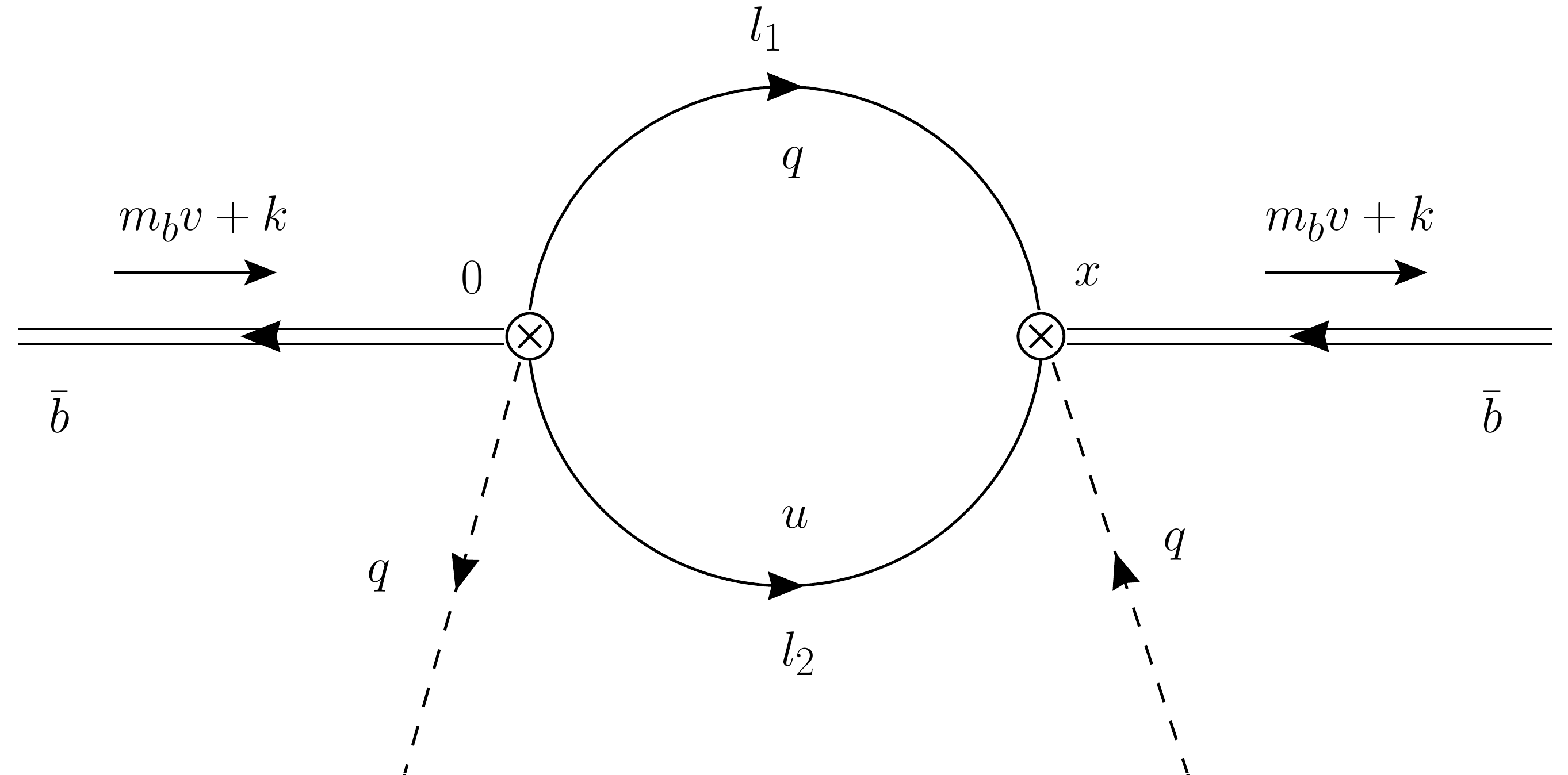} 
\caption{The diagram of $T_{ba}$. The initial and final $B$ mesons are replaced byt free $\bar b$ quarks with momentum $p_b=m_b v+k$. The two crossed dots denote ${\bar {\cal O}}_{(q)}^{\dagger}(x)$ and ${\bar {\cal O}}^I_{(uq)}(0)$ respectively.}
\label{fig:inclusivebuqpsi} 
\end{center}
\end{figure}

We firstly consider the type-I model. The $T_{ba}$ is calculated by HQE with the expansion for the power of $1/m_b$.  Using the explicit form of ${\bar {\cal O}}_{(uq)}^I$  
\begin{align}
{\bar {\cal O}}^I_{(uq)}=-i\epsilon_{ijk}({\bar b}_R^i u_R^{c,j}){\bar q}_R^{k},~~~
{\bar {\cal O}}^{I \dagger}_{(uq)} &=-i\epsilon_{ijk}({\bar u}_R^{c,i} b_R^{j})\gamma^0 q_R^{k}
\end{align}
and free quark propagators, one can obtain  
\begin{align}
T_{ba}=&\frac{i}{m_B}\int d^4x e^{-i q\cdot x} \int \frac{d^4 l_1}{(2\pi)^4}\frac{d^4 l_2}{(2\pi)^4} e^{-i l_1\cdot x}e^{-i l_2\cdot x}\nonumber\\
&\times \left[\gamma^0 P_R \frac{i(\slashed l_1+m_q)}{l_1^2-m_q^2}P_L\right]_{ba} \langle B(p_B)|{\bar b}^i(0) \frac{i \slashed l_2}{l_2^2-m_u^2} P_R b^{i^{\prime}}(x)|B(p_B)\rangle.
\end{align}
To extract the perturbative part of the matrix element above, one can temporary replace the initial and final $B$ meson with free $\bar b$ quark, namely $|B(p_B)\rangle\to |p_b\rangle$ with $p_b=m_b v+k$ and $k$ is of order $\Lambda_{QCD}$. Accordingly we can do the replacement
\begin{align}
\langle p_b|{\bar b}^i(0) \slashed l_2 P_R b^{i^{\prime}}(x)|p_b\rangle \to -e^{i p_b \cdot x}\ {\bar b}^i(p_b) \slashed l_2 P_R b^{i}(p_b) \to e^{i p_b \cdot x}\ \langle B(p_B)|{\bar b}(0) \slashed l_2 P_R b(0) |B(p_B)\rangle,
\end{align}
where $b(p_b)$ denotes the $\bar b$ quark spinor. In the last step the external states are transformed back to $B$ meson. Now the diagram of the correlation function $T$ is shown by Fig.\ref{fig:inclusivebuqpsi}, where the two crossed dots denote ${\bar {\cal O}}_{(q)}^{\dagger}(x)$ and ${\bar {\cal O}}_{(q)}(0)$ respectively. The $W_{ba}$ can be calculated by extracting the discontinuity part of $T_{ba}$ using cutting rules, namely all the internal quark lines in Fig.\ref{fig:inclusivebuqpsi} are set on-shell: $1/(l_1^2-m_q^2)\to (-2\pi i)\delta(l_1^2-m_q^2)$, $1/(l_2^2-m_u^2)\to (-2\pi i)\delta(l_1^2-m_u^2)$. 
Then we arrive at
\begin{align}
W_{ba}=&-\frac{1}{\pi}{\rm Im}\ T_{ba}=-\frac{1}{2\pi i}{\rm Disc}\ T_{ba}\nonumber\\
=&-\frac{(2\pi)^3}{m_B}\left\{A_{\rm 2bd}[(Q+k)^2,m_q^2,m_u^2](Q+k)^{\mu}(Q+k)^{\nu}+B_{\rm 2bd}[(Q+k)^2,m_q^2,m_u^2] (Q+k)^2 g^{\mu\nu}\right\}\nonumber\\
&\times \left[\gamma^0\gamma_{\mu} P_L\right]_{ba}\langle B(p_B)|{\bar b}(0) \gamma_{\nu} P_R b(0) |B(p_B)\rangle,\label{eq:Wba2bdy}
\end{align}
where $Q=m_b v-q$.  The $A_{\rm 2bd}, B_{\rm 2bd}$ are the two scalar functions of the rank-2 two-body phase space integration, which is generally defined as: 
\begin{align}
&\int \frac{d^4 l_1}{(2\pi)^3}\frac{d^4 l_2}{(2\pi)^3}\delta^4(P-l_1-l_2)\delta(l_1^2-m_1^2)\delta(l_2^2-m_2^2)l_1^{\mu}l_2^{\nu}\nonumber\\
=& A_{\rm 2bd}[P^2,m_1^2,m_2^2]P^{\mu}P^{\nu}+B_{\rm 2bd}[P^2,m_1^2,m_2^2] P^2 g^{\mu\nu}.
\end{align}
The explicit expression of $A_{\rm 2bd}$ and  $B_{\rm 2bd}$ are given in the Appendix \ref{2bdInt}.

The $1/m_b$ expansion is equivalent to the expansion in terms of the small momentum $k$. At ${\cal O}(k^0)$ all the $k$ s in Eq.~(\ref{eq:Wba2bdy}) vanishes and the $b$ quark field are replaced by the effective one $b_v$.  The axial-vector matrix element in Eq.~(\ref{eq:Wba2bdy}) vanishes due to the parity. The vector matrix element can be calculated straightforwardly  as
\begin{align}
\langle B(p_B)|{\bar b}_v(0) \gamma_{\nu} b_v(0) |B(p_B)\rangle=-2 m_B v_{\mu}
\end{align}
since ${\bar b}(0) \gamma_{\nu} b(0)$ is the conserved $b$ quark number current and the $b$ quark number of $B$ meson is  $-1$. In terms of the $k$ expansion, the lowest order of $W_{ba}$ reads as
\begin{align}
W_{ba}^{k^0}=&(2\pi)^3\left\{A_{\rm 2bd}[Q^2,m_q^2,m_u^2]Q^{\mu}v\cdot Q+B_{\rm 2bd}[Q^2,m_q^2,m_u^2] Q^2 v^{\mu}\right\}\left[\gamma^0\gamma_{\mu} P_L\right]_{ba},
\end{align} 
where $v\cdot Q=E$ and $ Q^2=m_b^2+m_{\psi}^2-2 m_b E$. The explicit expression of $A_{1,2}$ at ${\cal O}(k^0)$ are given in the Appendix \ref{A1A2expr}.

The ${\cal O}(k^1)$ contribution to $W_{ba}$ comes from the terms linear in $k$ in Eq.~(\ref{eq:Wba2bdy}). The procedure to extract the perturbative part by temporarily changing the external $B$ mesons to free $\bar b$ quark is almost the same as that at ${\cal O}(k^0)$. However, now the non-perturbative matrix element becomes $\langle B(p_B)|{\bar b} \gamma_{\nu}k_{\rho} P_R b |B(p_B)\rangle$, which can be written as $\langle B(p_B)|{\bar b} \gamma_{\nu}(iD_{\rho}-m_b v_{\rho}) P_R b |B(p_B)\rangle$ if transferred to coordinate space. Note that the $\gamma_5$ term vanishes again due to the reason of parity conservation.  Changing the $b$ quark field into the heavy quark field in HQET, one arrives at
\begin{align}
& \frac{1}{2}\langle B(p_B)|{\bar b} \gamma_{\nu}(iD_{\rho}-m_b v_{\rho}) b|B(p_B)\rangle\nonumber\\
=& \frac{1}{2}\langle B(p_B)|{\bar b}_v \gamma_{\nu} iD_{\rho} b_v|B(p_B)\rangle+\frac{i}{2}\int d^4 x  \langle B(p_B)|{\rm T} \left\{{\bar b}_v \gamma_{\nu}iD_{\rho} b_v(0) {\cal L}_1(x)\right\}|B(p_B)\rangle\nonumber\\
&+\frac{1}{2}\langle B(p_B)|{\bar b}_v \frac{-i \slashed{\overleftarrow D}}{2m_b}\gamma_{\nu}iD_{\rho} b_v|B(p_B)\rangle+\frac{1}{2}\langle B(p_B)|{\bar b}_v  \gamma_{\nu}iD_{\rho}\frac{i \slashed{\overrightarrow D}}{2m_b}b_v|B(p_B)\rangle\label{eq:ok1replacement},
\end{align}
where 
\begin{align}
{\cal L}_1=-{\bar b}_v \frac{D^2}{2 m_b}b_v-{\bar b}_v \frac{g}{4 m_b}G_{\alpha\beta}\sigma^{\alpha\beta}b_v
\end{align}
is the ${\cal O}(1/m_b)$ interaction term of the HQET Lagrangian. The matrix element of the first term in Eq.~(\ref{eq:ok1replacement}) vanishes because of the equation of motion,  while the second term in Eq.~(\ref{eq:ok1replacement}) can be parameterized as \cite{Manohar:1993qn}:
\begin{align}
\langle B(p_B)|\frac{i}{2}\int d^4 x {\rm T} \left\{{\bar b}_v \gamma_{\nu}iD_{\rho} b_v(0) {\cal L}_1(x)\right\}|B(p_B)\rangle=\frac{1}{2} m_B A\  v_{\nu}v_{\rho}\label{eq:defineA}
\end{align}
with 
\begin{align}
A=-\langle B(v)|{\cal L}_1(0)|B(v)\rangle=-\frac{\lambda_1}{m_b}-\frac{3\lambda_2}{m_b},
\end{align}
where ${\sqrt m_B}|B(v)\rangle=|B(p_B)\rangle$. The matrix element of the last two terms in Eq.~(\ref{eq:ok1replacement}) can be parameterized as
\begin{align}
\frac{1}{2}\langle B(p_B)|{\bar b}_v \frac{-i \slashed{\overleftarrow D}}{2m_b}\gamma_{\nu}iD_{\rho} b_v+{\bar b}_v  \gamma_{\nu}iD_{\rho}\frac{i \slashed{\overrightarrow D}}{2m_b}b_v|B(p_B)\rangle=m_B\frac{Y-Z}{4 m_b}(g_{\nu\rho}-v_{\nu}v_{\rho}),\label{eq:defineY-Z}
\end{align}
where $Y=(2/3)\lambda_1, Z=-4\lambda_2$ \cite{Manohar:1993qn}. Using the non-perturbative matrix elements defined in Eq.~(\ref{eq:defineA}) and Eq.~(\ref{eq:defineY-Z}), we obtain the ${\cal O}(k^1)$ contribution to $W_{ba}$ as
\begin{align}
W_{ba}^{k^1}=&-(2\pi)^3\big\{A_{\rm 2bd}[Q^2,m_q^2,0](Q^{\mu}g^{\nu\rho}+ Q^{\nu} g^{\mu\rho})+A_{\rm 2bd}[Q^2,m_q^2,0]^{(1)}2Q^{\rho}Q^{\mu}Q^{\nu}\nonumber\\
&+B_{\rm 2bd}[Q^2,m_q^2,0]2Q^{\rho}g^{\mu\nu}+B_{\rm 2bd}[Q^2,m_q^2,0]^{(1)}2Q^{\rho}Q^2 g^{\mu\nu}\big\}\nonumber\\
&\times \left[\gamma^0\gamma_{\mu} P_L\right]_{ba}\left\{\frac{1}{2}A v_{\nu}v_{\rho}+\frac{Y-Z}{4 m_b}(g_{\nu\rho}-v_{\nu}v_{\rho})\right\}.
\end{align}

Similarly, the ${\cal O}(k^2)$ contribution to $W_{ba}$ comes from:
\begin{align}
W_{ba}^{k^2}=&-\frac{(2\pi)^3}{m_B}\Big\{A_{\rm 2bd}[Q^2,m_q^2,0]g^{\mu\rho}g^{\nu\sigma}+A^{(1)}_{\rm 2bd}[Q^2,m_q^2,0]2Q^{\rho}(Q^{\mu}g^{\nu\sigma}+Q^{\nu}g^{\mu\sigma})\nonumber\\
&+\left[g^{\rho\sigma}A^{(1)}_{\rm 2bd}[Q^2,m_q^2,0]+2Q^{\rho}Q^{\sigma}A^{(2)}_{\rm 2bd}[Q^2,m_q^2,0]\right]Q^{\mu}Q^{\nu}+B_{\rm 2bd}[Q^2,m_q^2,0]g^{\mu\nu}g^{\rho\sigma}\nonumber\\
&+B^{(1)}_{\rm 2bd}[Q^2,m_q^2,0]4Q^{\rho}Q^{\sigma}g^{\mu\nu}+\left[g^{\rho\sigma}B^{(1)}_{\rm 2bd}[Q^2,m_q^2,0]+2Q^{\rho}Q^{\sigma}B^{(2)}_{\rm 2bd}[Q^2,m_q^2,0]\right]Q^2g^{\mu\nu}\Big\}\nonumber\\
&\times \left[\gamma^0\gamma_{\mu} P_L\right]_{ba}\langle B(p_B)|{\bar b}(0) \gamma_{\nu}k_{\rho}k_{\sigma} P_R b(0) |B(p_B)\rangle,
\end{align}
where $A^{(n)}_{\rm 2bd}=\partial^n/(\partial Q^2)^n A_{\rm 2bd}$ and similar for $B^{(n)}_{\rm 2bd}$. Each $k$ in the matrix element above is replaced by $iD -mv$ when transferred to coordinate space. Transforming the $b$ into $b_v$ and using the results given in Ref.\cite{Manohar:1993qn}, we have
\begin{align}
\langle B(p_B)|{\bar b} \gamma_{\nu}k_{\rho}k_{\sigma} P_R b |B(p_B)\rangle=\frac{1}{2}\langle B(p_B)|{\bar b}_v v_{\nu}(g_{\rho\sigma}-v_{\rho}v_{\sigma}) b_v |B(p_B)\rangle=\frac{1}{2}m_B Y v_{\nu} (g_{\rho\sigma}-v_{\rho}v_{\sigma}).
\end{align}
The explicit expressions of $A_{1,2}$ at ${\cal O}(k^1)$ and ${\cal O}(k^2)$ are given in  the Appendix \ref{A1A2expr}.

\begin{figure}
\begin{center}
\includegraphics[width=1.0\columnwidth]{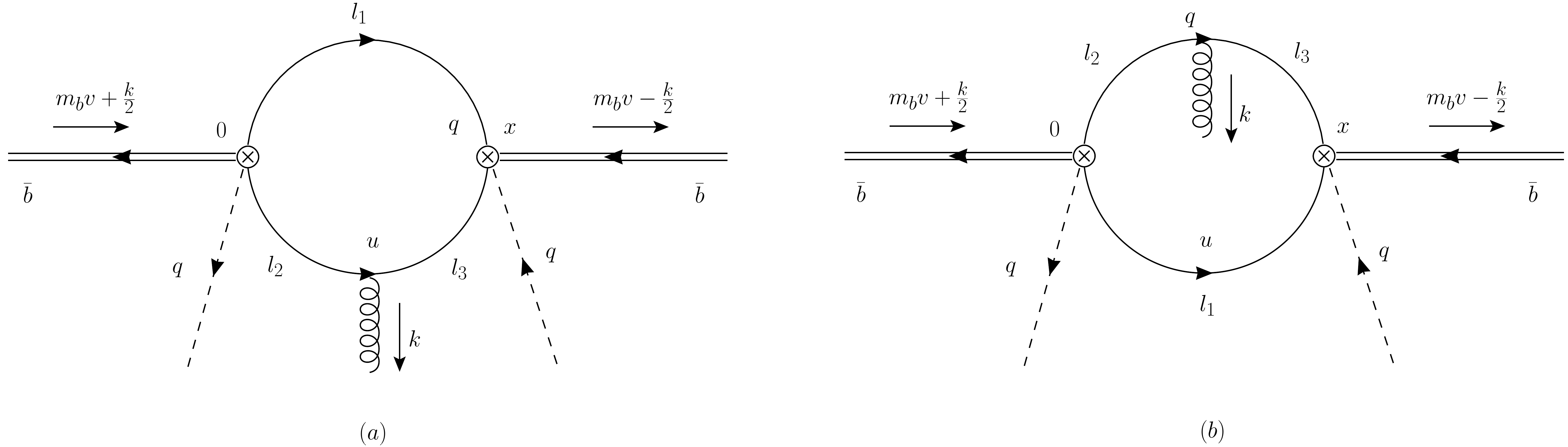} 
\caption{One gluon emission diagrams of $T_{ba}$. The incoming and outgoing $B$ mesons are replaced by free $\bar b$ states, with momentums being  $p_1=m_b v+k/2$ and $p_2=m_b v-k/2$ respectively.}
\label{fig:inclusivebugu} 
\end{center}
\end{figure}
Up to now we have only considered the case of free quark propagation when calculating the $T_{ba}$ as shown in  Fig.\ref{fig:inclusivebuqpsi}. When considering the interaction of the internal quarks and the background gluon fields, one has to calculated the one gluon emission diagrams as shown in Fig. \ref{fig:inclusivebugu}. Here 
the external $B$ mesons are also replaced by free $\bar b$ states. We have set the incoming and outgoing $b$ quark momentums as $p_1=m_b v+k/2$ and $p_2=m_b v-k/2$ respectively.  Here we take the $u$ quark emission as an example, the corresponding $T_{ba}$ tensor is:
\begin{align}
T_{ba}^{ug}=&\frac{ig}{2m_B}t_{ij}^a \epsilon_{\mu}^{a *}(k) \frac{1}{(2\pi)^4}\int d^4 l_1 \frac{[\gamma^0 \gamma_{\rho}P_L]_{ba}}{(l_1^2-m_s^2)\left[\left(Q-l_1+\frac{k}{2}\right)^2-m_u^2\right]\left[\left(Q-l_1-\frac{k}{2}\right)^2-m_u^2\right]}\nonumber\\
&\times \left\{\left(Q-l_1+\frac{k}{2}\right)_{\alpha}\left(Q-l_1-\frac{k}{2}\right)_{\beta}{\bar b}^i(p_2) \gamma^{\alpha}\gamma^{\mu}\gamma^{\beta}P_R b^j(p_1)+m_u^2 {\bar b}^i(p_2)\gamma^{\mu} P_R b^j(p_1)\right\}.
\end{align}
The emitted gluon has momentum $k$ and note that now the ${\cal O}(k^1)$ term in the denominator vanishes. The ${\cal O}(k^1)$ contribution to $T_{ba}^{ug}$ is
\begin{align}
T_{ba}^{ug,k^1}
=&\frac{ig}{2m_B}t_{ij}^a \epsilon_{\mu}^{a *}(k) \frac{1}{2(2\pi)^4}\frac{\partial}{\partial M^2}\int d^4 l_1 d^4 l_2 \delta^4(Q-l_1-l_2) \frac{1}{(l_1^2-m_s^2)(l_2^2-M^2)}\nonumber\\
&\times \left(k_{\alpha}l_{1\rho}l_{2\beta}-k_{\beta}l_{1\rho}l_{2\alpha}\right){\bar b}^i(p_2) \gamma^{\alpha}\gamma^{\mu}\gamma^{\beta}P_R b^j(p_1)[\gamma^0 \gamma_{\rho}P_L]_{ba},
\end{align}
where we have used the trick $1/(l_2^2-m_2^2)^2\to \partial_{M^2} \{1/(l_2^2-M^2)\}|_{M^2= m_2^2}$. The corresponding $W_{ba}$ can still be extracted by cutting rules, and thus we obtain
\begin{align}
W_{ba}^{ug,k^1}=&\frac{i g(2\pi)^3}{4m_B}t_{ij}^a \epsilon_{\mu}^{a *}(k)\partial_{M^2}\left[A_{\rm 2bd}[Q^2,m_q^2,M^2]Q^{\rho}Q_{\sigma}\epsilon^{\mu\nu\alpha\sigma}-B_{\rm 2bd}[Q^2,m_q^2,M^2]Q^2\epsilon^{\mu\nu\rho\alpha}\right]\nonumber\\
&\times \langle B(p_B)|{\bar b}^i(0) \gamma_{\nu}k_{\alpha}(1+\gamma_5) b^j(0)|B(p_B)\rangle [\gamma^0 \gamma_{\rho}P_L]_{ba}.
\end{align}
The combination of $k$ and $\epsilon^{a *}(k)$ can be replaced by the gluon tensor field when transferred to coordinate space. Explicitly, we can do the  replacement $k_{\alpha}\epsilon_{\mu}^{a*}t_{ij}^a\to (-i/2)G_{\alpha\mu}^a t_{ij}^a$ and $b\to b_v$, and also note that $\langle B(p_B)|{\bar b}_{v}^i(0) \gamma_{\nu}g G_{\alpha\mu} b_v^j(0) | B(p_B)\rangle=0$ and 
\begin{align}
\langle B(p_B)|{\bar b}_{v}^i(0) \gamma_{\nu}g G_{\alpha\mu}\gamma_5 b_v^j(0) | B(p_B)\rangle=m_B N\epsilon_{\alpha\mu\nu\kappa}v^{\kappa},
\end{align}
then we obtain the $W_{ba}$ for $u$ and $q$ emission as
\begin{align}
W_{ba}^{ug,k^1}=&-\frac{3}{4}(2\pi)^3 N \partial_{M^2}\Big\{A_{\rm 2bd}[Q^2,m_q^2,M^2](v\cdot Q)Q^{\rho}\nonumber\\
&+B_{\rm 2bd}[Q^2,m_q^2,M^2]Q^2 v^{\rho}\Big\} [\gamma^0 \gamma_{\rho}P_L]_{ba}\Big |_{M^2\to m_u^2}\nonumber\\
W_{ba}^{qg,k^1}=&-\frac{(2\pi)^3}{2} N \partial_{M^2}\Big\{3 B_{\rm 2bd}[Q^2,M^2,m_u^2]Q^2 v^{\rho}\nonumber\\
&-A_{\rm 2bd}[Q^2,M^2,m_u^2](Q \cdot v\  Q^{\rho}-Q^2 v^{\rho})\Big\} [\gamma^0 \gamma_{\rho}P_L]_{ba}\Big |_{M^2\to m_q^2}.
\end{align}
The corresponding explicit expressions of their contribution to $A_{1,2}$  are given in the Appendix \ref{A1A2expr}.

\subsection{Heavy quark expansion in the Type-II model}
In this section we will consider the type-II model. Now the HQE calculation of $T_{ba}$ is almost the same as that in the type-I model.  Using the explicit form of ${\bar {\cal O}}_{(uq)}^{II}$  
\begin{align}
{\bar {\cal O}}^{II}_{(uq)}=-i\epsilon_{ijk}({\bar q}_R^i u_R^{c,j}){\bar b}_R^{k},~~~
{\bar {\cal O}}^{II \dagger}_{(uq)} &=-i\epsilon_{ijk}({\bar u}_R^{c,i} q_R^{j})\gamma^0 b_R^{k},
\end{align}
and extracting the imaginary part of $T_{ba}$ as shown in Fig.\ref{fig:inclusivebuqpsi}, we can obtain the corresponding $W_{ba}$ through $W=-(1/\pi){\rm Im}T$: 
\begin{align}
W_{ba}=\frac{(2\pi)^4}{\pi m_B} C_{2bd}[(Q+k)^2,m_q^2,m_u^2](Q+k)^2\langle B(p_B)|[\gamma^0 P_R b^i(0)]_b [{\bar b}^i(0)P_L]_a|B(p_B)\rangle,
\end{align}
where $C_{2bd}=A_{2bd}+4B_{2bd}$. Note that now the spinor structure of the matrix element above is different from that of Eq.~(\ref{eq:Wba2bdy}), and it seems not straightforward to read out the $A_{1,2}$ defined in Eq.~(\ref{eq:parameterizeW}). However, instead one can use the following trick:
\begin{align}
{\rm tr}[\gamma_{\mu}\gamma^0 W]=2A_1\frac{q_{\mu}}{m_B}+2A_2 v_{\mu}.
\end{align}
to extract $A_{1,2}$. At ${\cal O}(k^0)$, ${\cal O}(k^1)$ and ${\cal O}(k^2)$ we have
\begin{align}
{\rm tr}[\gamma_{\mu}\gamma^0 W^{k^0}]=&\frac{(2\pi)^4}{\pi}C_{2bd}[Q^2,m_q^2,m_u^2]Q^2 v_{\mu},\nonumber\\
{\rm tr}[\gamma_{\mu}\gamma^0 W^{k^1}]=&-\frac{(2\pi)^4}{\pi}\left(C_{2bd}[Q^2,m_q^2,m_u^2]+C^{(1)}_{2bd}[Q^2,m_q^2,m_u^2]\right)Q^{\rho}\nonumber\\
&\times \left[A v\cdot Q v_{\mu}+\frac{Y-Z}{2m_b}(Q_{\mu}-v\cdot Q v_{\mu})\right],\nonumber\\
{\rm tr}[\gamma_{\mu}\gamma^0 W^{k^2}]=&-(2\pi)^3 Y v_{\mu}\left[3C_{2bd}[Q^2,m_q^2,m_u^2]+2C^{(2)}_{2bd}[Q^2,m_q^2,m_u^2](Q^2)^2\right.\nonumber\\
&\left.+Q^2 \left(7C^{(1)}_{2bd}[Q^2,m_q^2,m_u^2]-2C^{(2)}_{2bd}[Q^2,m_q^2,m_u^2](v\cdot Q)^2\right)\right.\nonumber\\
&\left.-4C^{(1)}_{2bd}[Q^2,m_q^2,m_u^2](v\cdot Q)^2\right].
\end{align}

On the other hand, it can be found that in the Type-II model the ${\cal O}(k^1)$ contribution from the one gluon emission diagrams as shown in Fig. \ref{fig:inclusivebugu} vanishes. The explicit expression of $A_{1,2}$ at ${\cal O}(k^0)$, ${\cal O}(k^1)$ and ${\cal O}(k^2)$ are given in the Appendix \ref{A1A2exprTypeII}, which are proportional to $m_q^2$. Therefore, in the Type-II model the decay width of $B\to X_{ud}/X_{cd}\psi$ vanishes in the chiral limit $m_{u,d}=0$, and the decay width of $B\to X_{us}/X_{cs}\psi$ is suppressed compared with that  in the Type-I model.

\section{Numerical results}

In this section, we will present the numerical results on the various of $B\to X_{uq}\psi$ branching fractions as functions of $\psi$ mass. The mass parameters are $m_B=5.28$ GeV, $m_s=87$ MeV, $m_c=1.0$ GeV, $m_b=4.47\pm 0.03$ GeV\cite{ParticleDataGroup:2022pth}, where the quark masses are chosen at $\mu=3$ GeV as that used in Ref.~\cite{Khodjamirian:2022vta}. The non-perturbative parameters $\lambda_{1,2}$ are related with the kinetic term $\mu_{\pi}^2$ and the chromo-magnetic term $\mu_{G}^2$ of $B$ meson as: $\lambda_1=-\mu_{\pi}^2=-0.414\pm 0.078$ GeV$^2$ and $\lambda_2=\mu_{G}^2/3=0.117\pm0.023$ GeV$^2$ respectively \cite{Gambino:2013rza,Uraltsev:2001ih}. The errors of $m_b$ and $\lambda_{1,2}$ will be used for estimating the uncertainty of the numerical results.

Before calculating the decay width by Eq.~(\ref{eq:diffWidth2}), one has to determine the integration range of $E$. Obviously, the lower bound of $E$ must be $m_{\psi}$. On the other hand, the upper bound of $E$ seems to be $E_{\rm upper}=[m_b^2+m_{\psi}^2-(m_q+m_u)^2]/2m_b$, which is reached when the invariant momentum square $Q^2$ flowing into the loop bubble as shown in Fig.\ref{fig:inclusivebuqpsi} and Fig.\ref{fig:inclusivebugu} becomes  $Q^2=(m_q+m_u)^2$. However, it can be found that the terms proportional to $\lambda_{1,2}$ in the results of $A_{1,2}$ contain end point singularities at $E=E_{\rm upper}$, which can be seen from the pole structures $1/[Q^2-(m_q+m_u)^2]^{(n)}$ of $A_{2bd}^{(1,2)}$ and $B_{2bd}^{(1,2)}$, with $n$ being $1/2$ or $3/2$. Note that although $A_{2bd}$ and $B_{2bd}$ also have such pole structures, they are actually finite in the limit $Q^2\to(m_q+m_u)^2$. 
The reason why this end point singularity emerges is due to the fact that HQE breaks down at this region, where single states or resonances dominate. Before the expansion of $k$, the $W_{ba}$ contain the terms like $1/[Q^2-(m_q+m_u)^2+2 Q\cdot k +k^2]^{(n)}$. When $Q^2-(m_q+m_u)^2$ is large, the expansion of $k$ is right. However, when $Q^2\sim (m_q+m_u)^2$, this expansion is forbidden. 

It should be noted that the final states $X_{uq}$ observed in the experiment are baryons, not the quarks. Practically, one has to sum the inclusive states $X_{uq}$ from the lowest baryon state ${\cal B}_{uq}$. For example, in terms of the $B_0\to X_{ud}\psi$ decay,  ${\cal B}_{uq}$ is a proton or neutron. Accordingly, the lower bounds of $Q^2$ should be set as the mass square of the corresponding lowest baryon state, namely $Q^2>m_{{\cal B}_{uq}}^2$ or equivalently $E_{\rm upper}=[m_b^2+m_{\psi}^2-m_{{\cal B}_{uq}}^2]/2m_b$, and thus the end point singularity is avoided. Here we  have omitted the contribution of the spectator quark in $B$ meson to $X_{uq}$, because the energy of $X_{uq}$ is mostly given by the heavy $\bar b$ quark. The lower bounds of $Q^2$ corresponding to various of $b\to uq \psi$ transitions are listed in Table.\ref{Tab:LowBoundQ2}. 
\begin{table}
  \caption{The lower bounds of $Q^2$ is set as the mass square of the lowest baryon state in the summation of $X_{uq}$: $Q^2=m_{{\cal B}_{uq}}^2$. The contribution of the spectator quark in $B$ meson to $X_{uq}$ is omitted since the energy of $X_{uq}$ is mostly given by the heavy $\bar b$ quark.}
\label{Tab:LowBoundQ2}
\begin{tabular}{|c|cccc|}
\hline 
\hline
 Decay & $\bar b \to u d \psi$ & $\bar b \to u s \psi$ & $\bar b \to c d \psi$ & $\bar b \to c s \psi$ \tabularnewline
\hline 
Lowest $X_{uq}$ & $p/n$ & $\Lambda$ & $\Lambda_c$ & $\Xi_c$ \tabularnewline
$m_{{\cal B}_{uq}} (\rm GeV)$ \cite{ParticleDataGroup:2022pth}  & $1.0$ & $1.115$ & $2.286$ & $2.468$ \tabularnewline
\hline 
\end{tabular}
\end{table}

\begin{figure}
\begin{center}
\includegraphics[width=0.45\columnwidth]{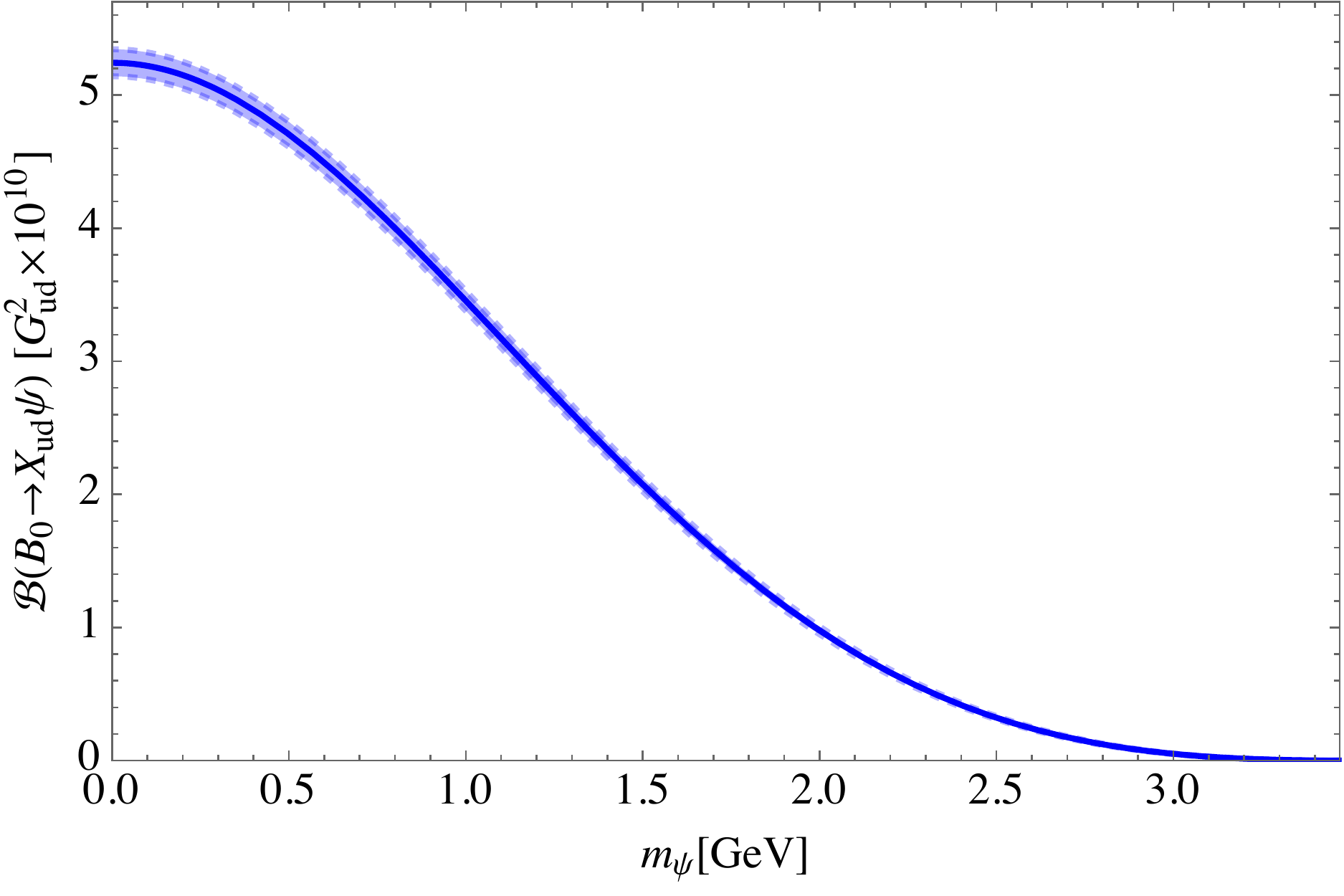} 
\includegraphics[width=0.45\columnwidth]{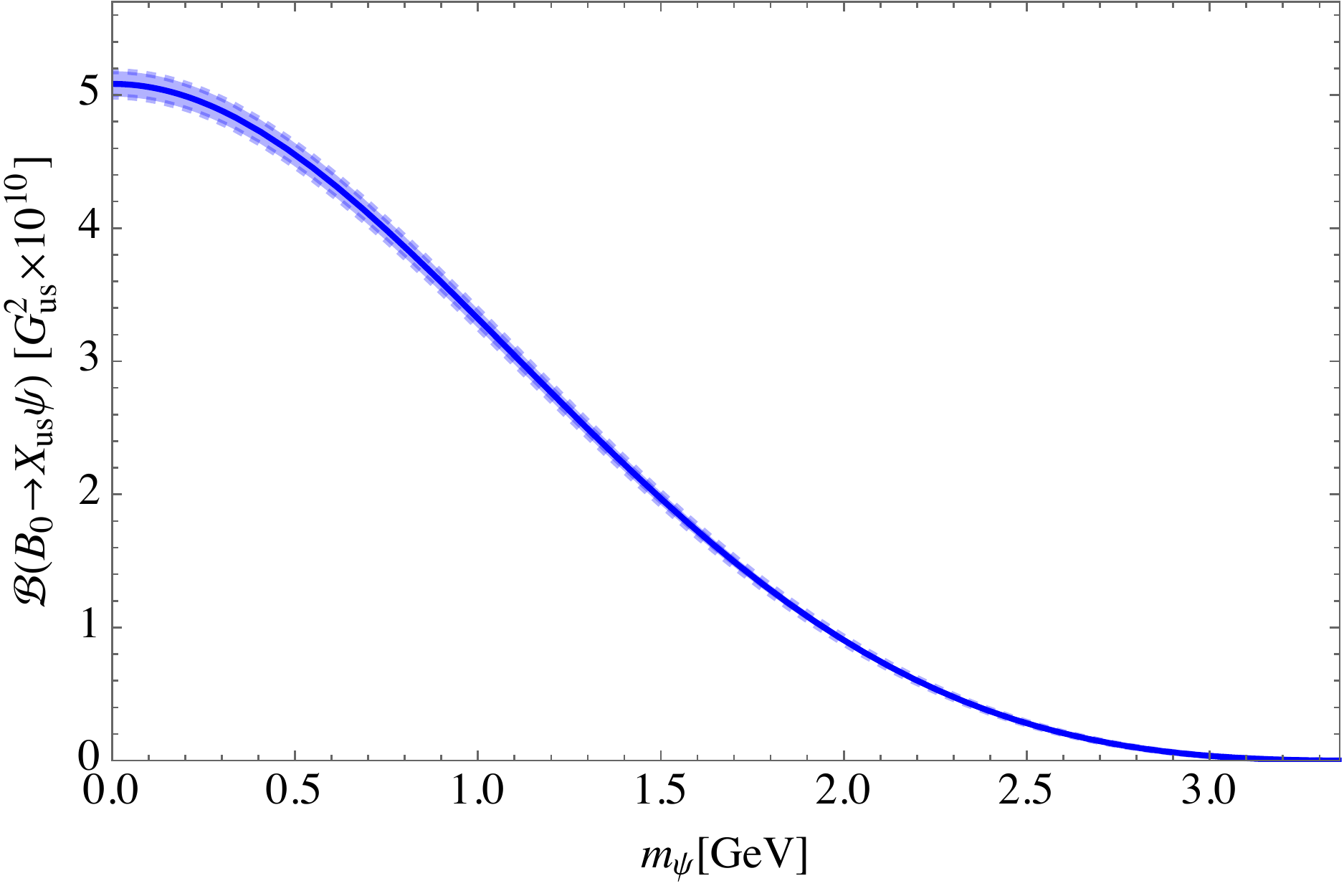} 
\includegraphics[width=0.45\columnwidth]{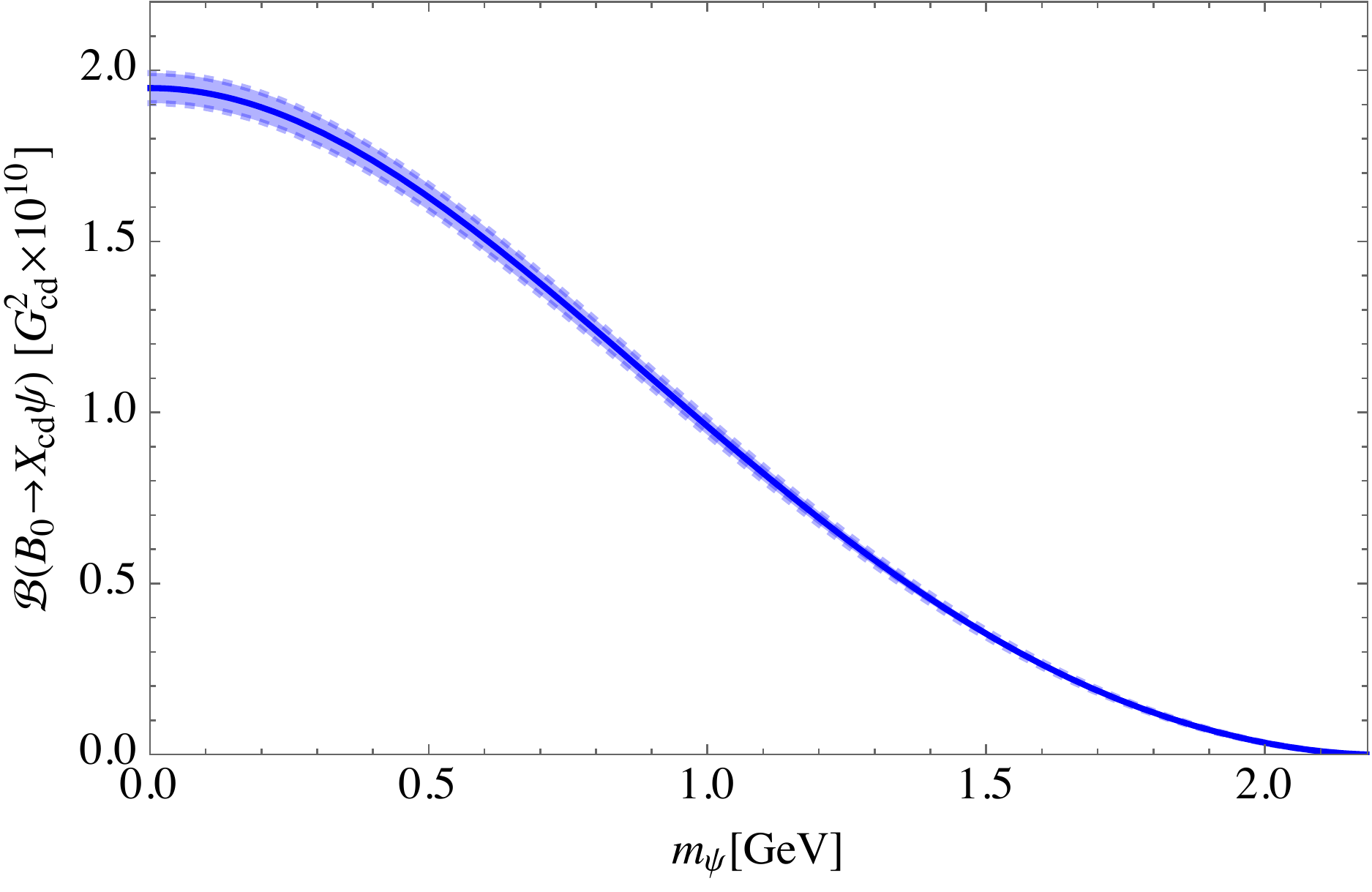} 
\includegraphics[width=0.45\columnwidth]{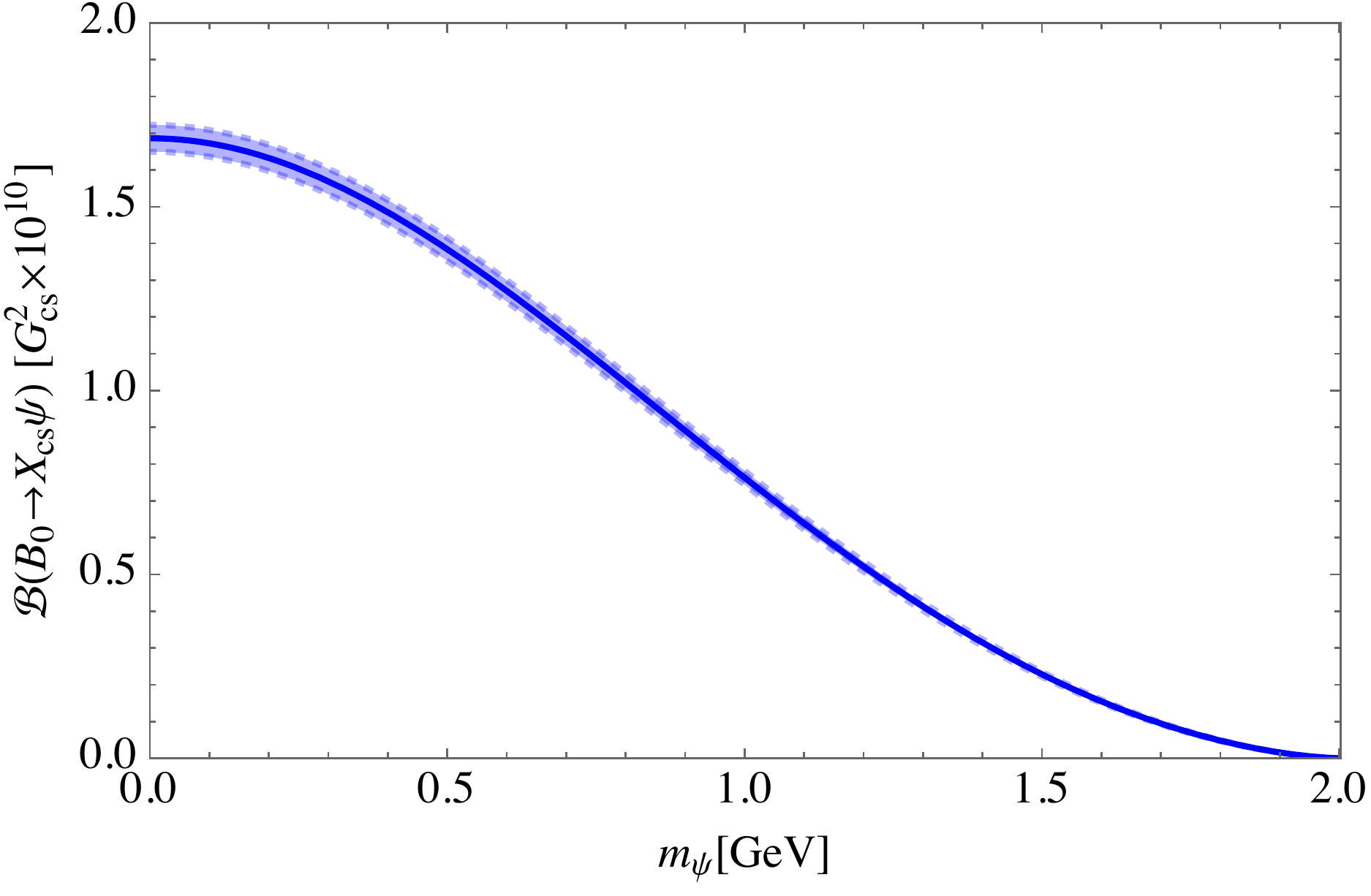} 
\caption{The type-I model branching fractions of $B_0\to X_{ud}\psi$, $B_0\to X_{us}\psi$, $B_0\to X_{cd}\psi$ and $B_0\to X_{cs}\psi$ as functions of $m_{\psi}$ in the  unit of $G_{uq}^2\times 10^{10}$. The band width shows the uncertainty coming from the uncertainties of $\lambda_{1,2}$ and $m_b$. The maximum $m_{\psi}$ is reached when $m_{\psi}=E_{\rm upper}$.}
\label{fig:B0toXuqpsiTypeI} 
\end{center}
\end{figure}
\begin{figure}
\begin{center}
\includegraphics[width=0.45\columnwidth]{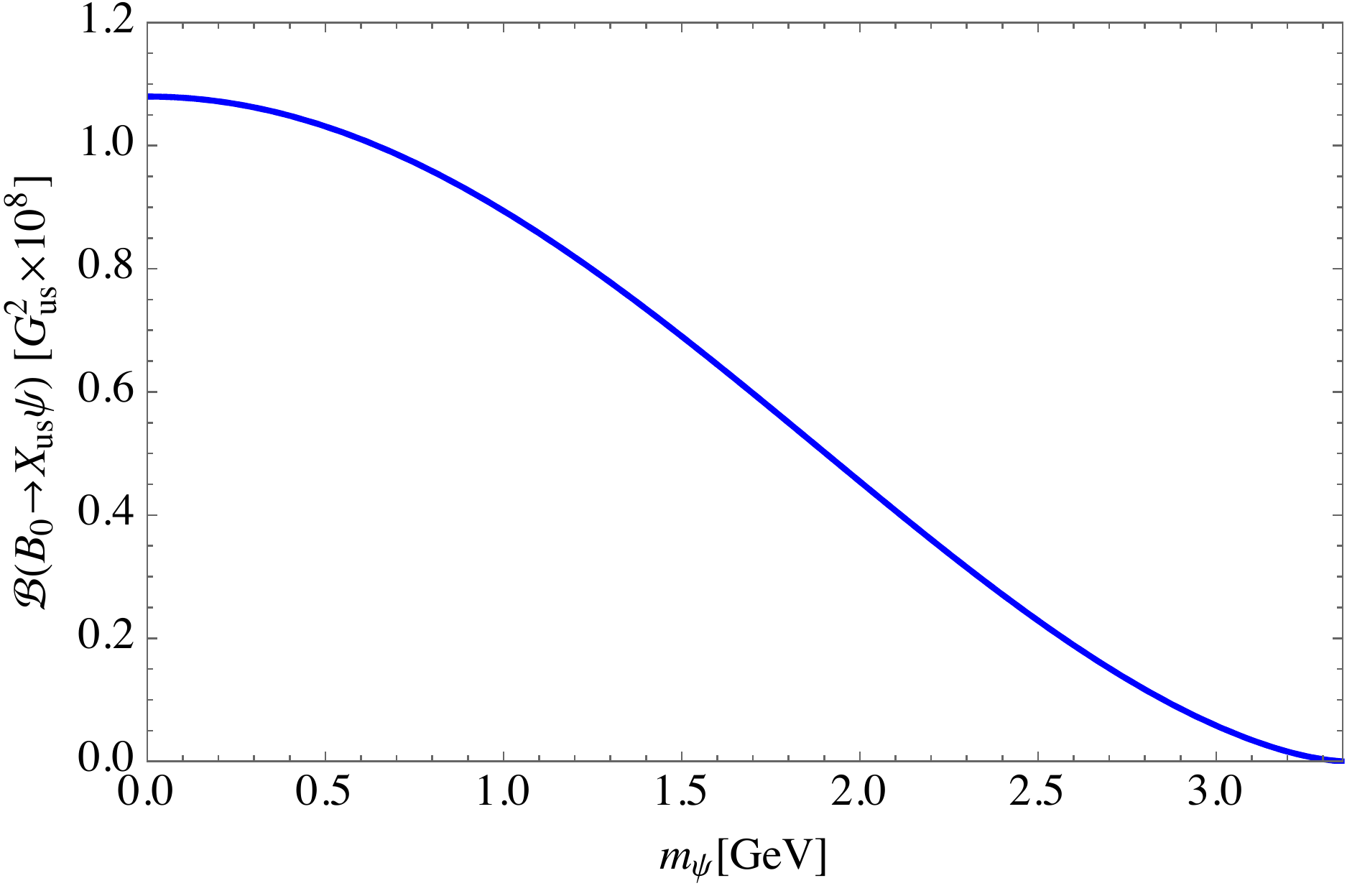} 
\includegraphics[width=0.45\columnwidth]{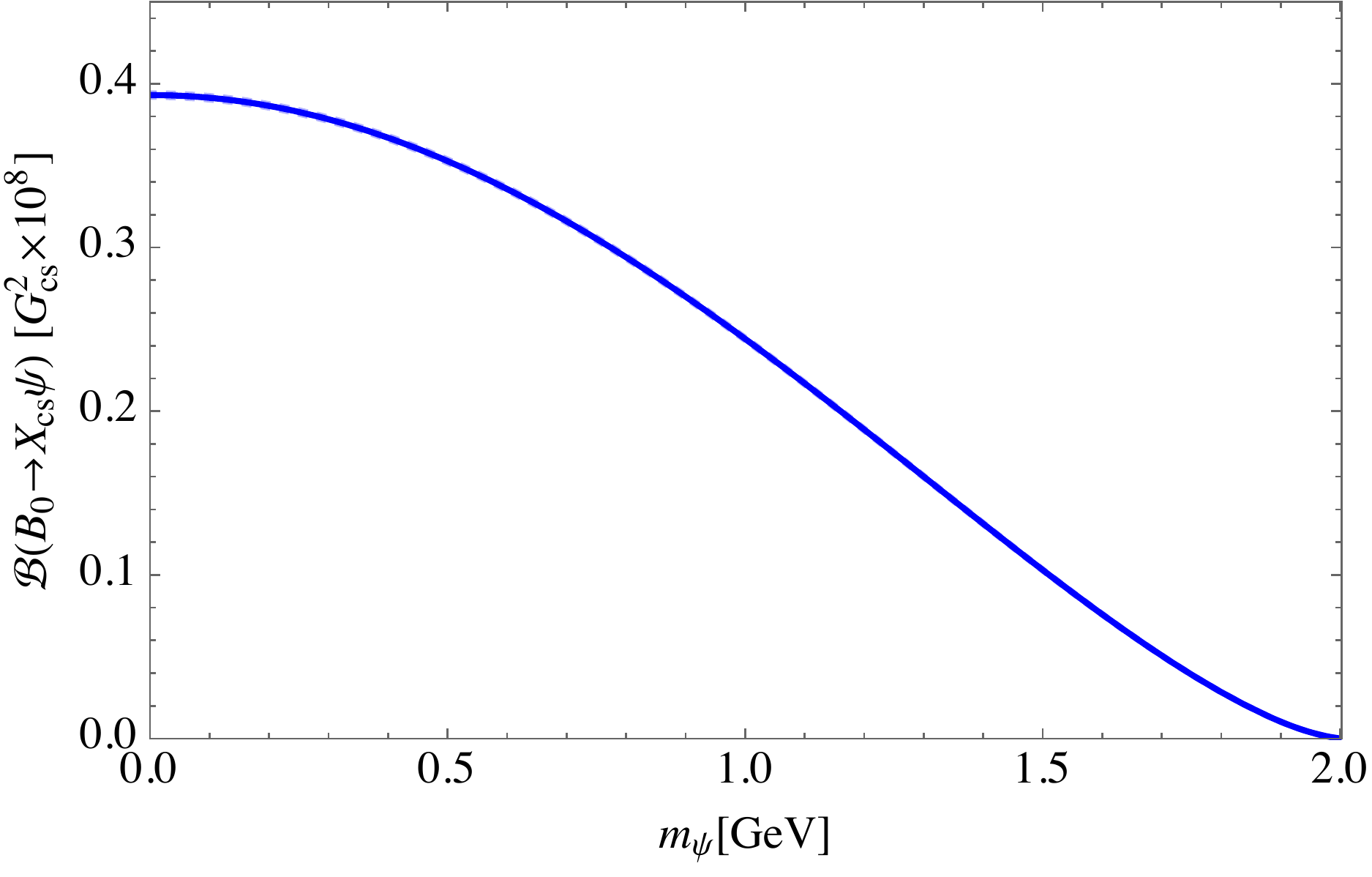} 
\caption{The type-II model branching fractions of $B_0\to X_{us}\psi$ and $B_0\to X_{cs}\psi$ as functions of $m_{\psi}$ in the unit of $G_{uq}^2\times 10^{8}$. The maximum $m_{\psi}$ is reached when $m_{\psi}=E_{\rm upper}$. The branching fractions of $B_0\to X_{ud}\psi$ and $B_0\to X_{cd}\psi$ vanish due to $m_u=m_d=0$. The uncertainty mainly comes from $\lambda_{1,2}$, which is tiny and can be ignored.}
\label{fig:B0toXuqpsiTypeII} 
\end{center}
\end{figure}
Now, integrating $E$ in the region $m_{\psi}<E<E_{\rm upper}$, and using the lifetime of $B_0$: $\tau_{B_0}=1.519\times 10^{-12}$ fs, we can obtain the branching fractions of $B_0\to X_{uq}\psi$ as functions of $m_{\psi}$. The branching fractions calculated in the type-I model are shown in Fig.~\ref{fig:B0toXuqpsiTypeI} in the unit of $G_{uq}^2\times 10^{10}$. The band width shows the uncertainty coming from the uncertainties of $\lambda_{1,2}$ and $m_b$.\footnote{The error is estimated by the formula: $\delta \Gamma=\sqrt{\left(\delta \lambda_1 \frac{\partial \Gamma}{\partial \lambda_1}\right)^2+\left(\delta \lambda_2 \frac{\partial \Gamma}{\partial \lambda_2}\right)^2+\left(\delta m_b \frac{\partial \Gamma}{\partial m_b}\right)^2}$, where $\delta \lambda_{1,2}$ and $\delta m_b$ are the errors of $\lambda_{1,2}$ and $m_b$ respectively.} In fact, the results are insensitive to the values of $\lambda_{1,2}$, and most of the uncertainties come from the $b$ quark mass since in the type-I model $A_{1,2}$ are proportional to $m_b^n$. The maximum $m_{\psi}$ is reached when $m_{\psi}=E_{\rm upper}$. The branching fractions calculated in the type-II model are shown in Fig.~\ref{fig:B0toXuqpsiTypeII} in the unit of $G_{uq}^2\times 10^{8}$. It should be mentioned that in the type-II model, the $A_{1,2}$ are proportional to $m_q$, and thus vanishes in the case of $B_0\to X_{ud}\psi$ and $B_0\to X_{cd}\psi$. In Fig.~\ref{fig:B0toXuqpsiTypeII} only the branching fractions of $B_0\to X_{us}\psi$ and $B_0\to X_{cs}\psi$ are presented. The uncertainty mainly comes from $\lambda_{1,2}$, which is tiny and can be ignored. Since the masses and lifetimes of $B^{\pm}$ and $B_s$ are similar to those of $B_0$, in this work we only present the branching fractions of $B_0$ decays and the decay branching fractions of $B^{\pm}$ and $B_s$ are assumed to be the same.

In addition,  instead of the complicated analytical expression given in the Appendix \ref{A1A2expr} and \ref{A1A2exprTypeII}, for practice we can parameterize the branching fraction curves shown in Fig.~\ref{fig:B0toXuqpsiTypeI} and Fig.~\ref{fig:B0toXuqpsiTypeII} by a simpler formula. Here we use the following polynomial form to fit the curves:
\begin{align}
{\cal B}(B_0\to X_{uq}\psi)=G_{uq}^2\times 10^{n} \times \sum_{i=0}^{7} a_{i} \left(\frac{m_{\psi}}{m_{B}}\right)^i,\label{eq:approxFormula}
\end{align}
where $n=10, 8$ for the case of type-I, II. $a_i$ have unit GeV$^4$ and are listed in Table. \ref{Tab:aiList}. It should be mentioned that the number of the polynomial terms given above is chosen arbitrarily, which is enough for parameterizing the curves perfectly.
\begin{table}
  \caption{The coefficients $a_i$ defined in the Eq.~(\ref{eq:approxFormula}) (in unit GeV$^4$). Due to the tiny error bar of the type-II branching fraction, here we ignore the errors of $a_i$ in the type-II case.}
\label{Tab:aiList}
\scalebox{0.85}{
\begin{tabular}{|c|c|c|c|c|c|c|c|c|}
\hline 
\hline
 Type-I & $a_0$ & $a_1$ & $a_2$ & $a_3$ & $a_4$ & $a_5$ & $a_6$ & $a_7$ \tabularnewline
\hline 
$\bar b \to u d \psi$ & $5.24\pm 0.1$ & $0.07\pm 0.0$ & $-68.76\pm 0.9$ & $61.5\pm 0.53$ & $313.58\pm 2.57$ & $-756.23\pm 3.66$ & $641.08\pm 1.24$ & $-198.06\pm 0.14$   \tabularnewline
$\bar b \to u s \psi$ & $5.08\pm 0.09$ & $0.07\pm 0.0$ & $-67.88\pm 0.88$ & $60.14\pm 0.54$ & $318.89\pm 2.44$ & $-771.56\pm 3.4$ & $659.65\pm 0.89$ & $-206.37\pm 0.35$    \tabularnewline
$\bar b \to c d \psi$ & $1.95\pm 0.04$ & $0.04\pm 0.0$ & $-42.35\pm 0.73$ & $49.82\pm 0.97$ & $245.86\pm 0.95$ & $-606.4\pm 12.73$ & $324.72\pm 35.48$ & $132.82\pm 30.68$    \tabularnewline
$\bar b \to c s \psi$ & $1.68\pm 0.04$ & $0.04\pm 0.0$ & $-40.54\pm 0.69$ & $54.05\pm 1.16$ & $194.55\pm 3.08$ & $-348.1\pm 23.81$ & $-332.342\pm 63.63$ & $777.545\pm 58.34$    \tabularnewline
\hline
Type-II & $a_0$ & $a_1$ & $a_2$ & $a_3$ & $a_4$ & $a_5$ & $a_6$ & $a_7$ \tabularnewline
\hline
$\bar b \to u s \psi$ & $1.08$ & $0.02$ & $-6.02$ & $5.98$ & $-22.19$ & $79.45$ & $-106.76$ & $53.08$   \tabularnewline
$\bar b \to c s \psi$ & $0.39$ & $0.02$ & $-5.7$ & $19.64$ & $-154.29$ & $732.86$ & $-1611.17$ & $1403.16$   \tabularnewline
\hline 
\end{tabular}}
\end{table}

\begin{figure}
\begin{center}
\includegraphics[width=0.45\columnwidth]{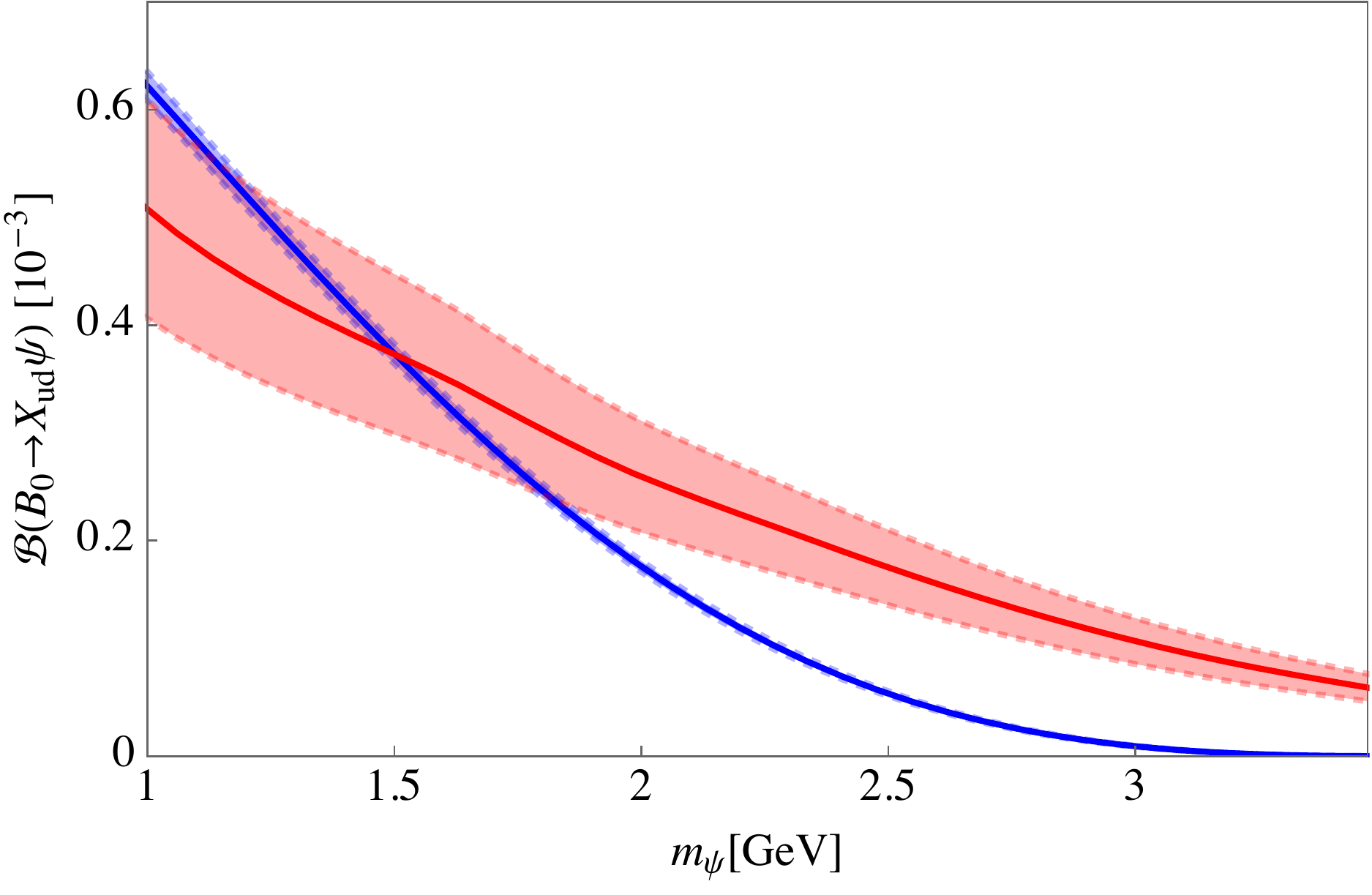} 
\includegraphics[width=0.45\columnwidth]{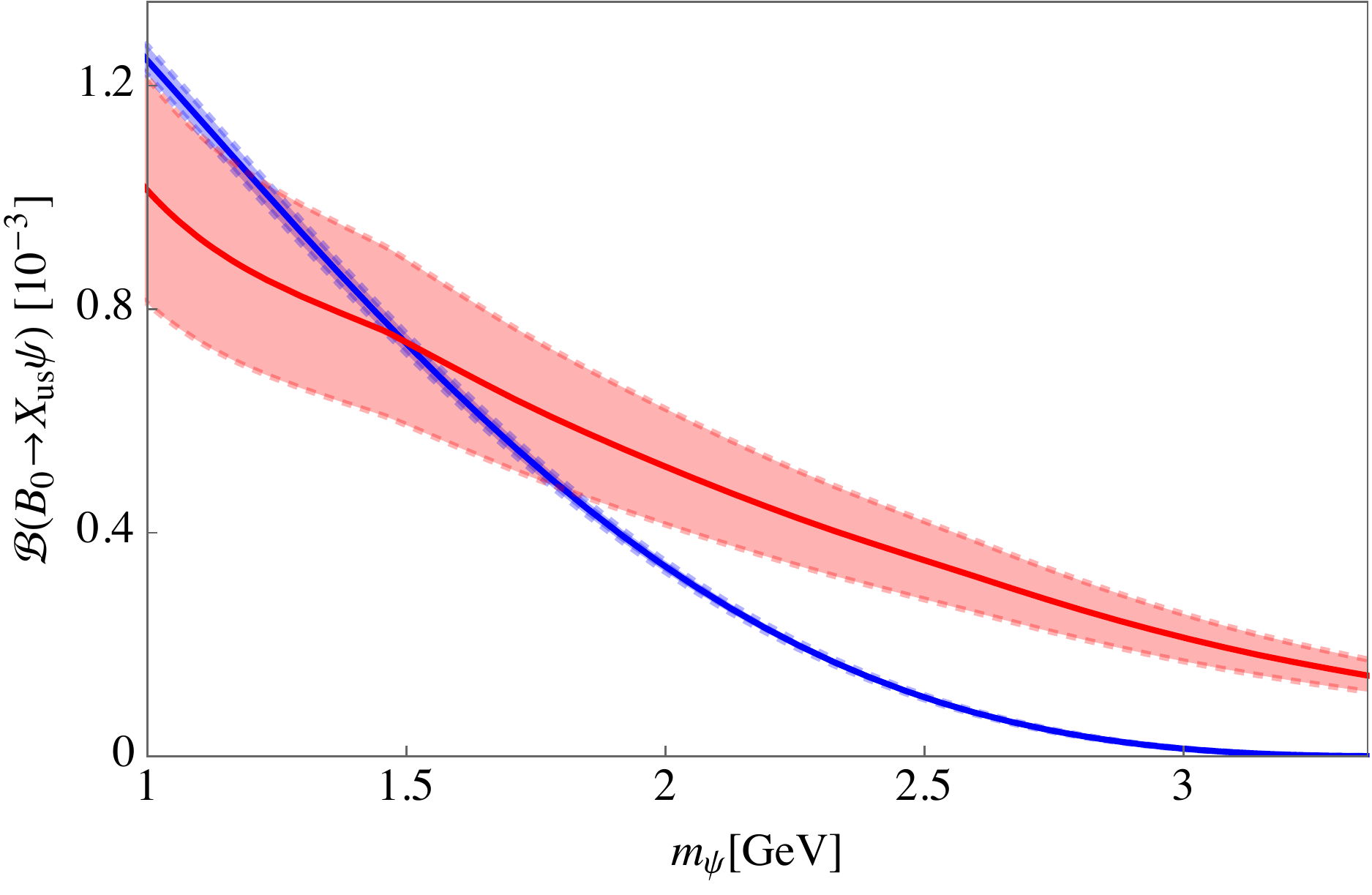} 
\includegraphics[width=0.45\columnwidth]{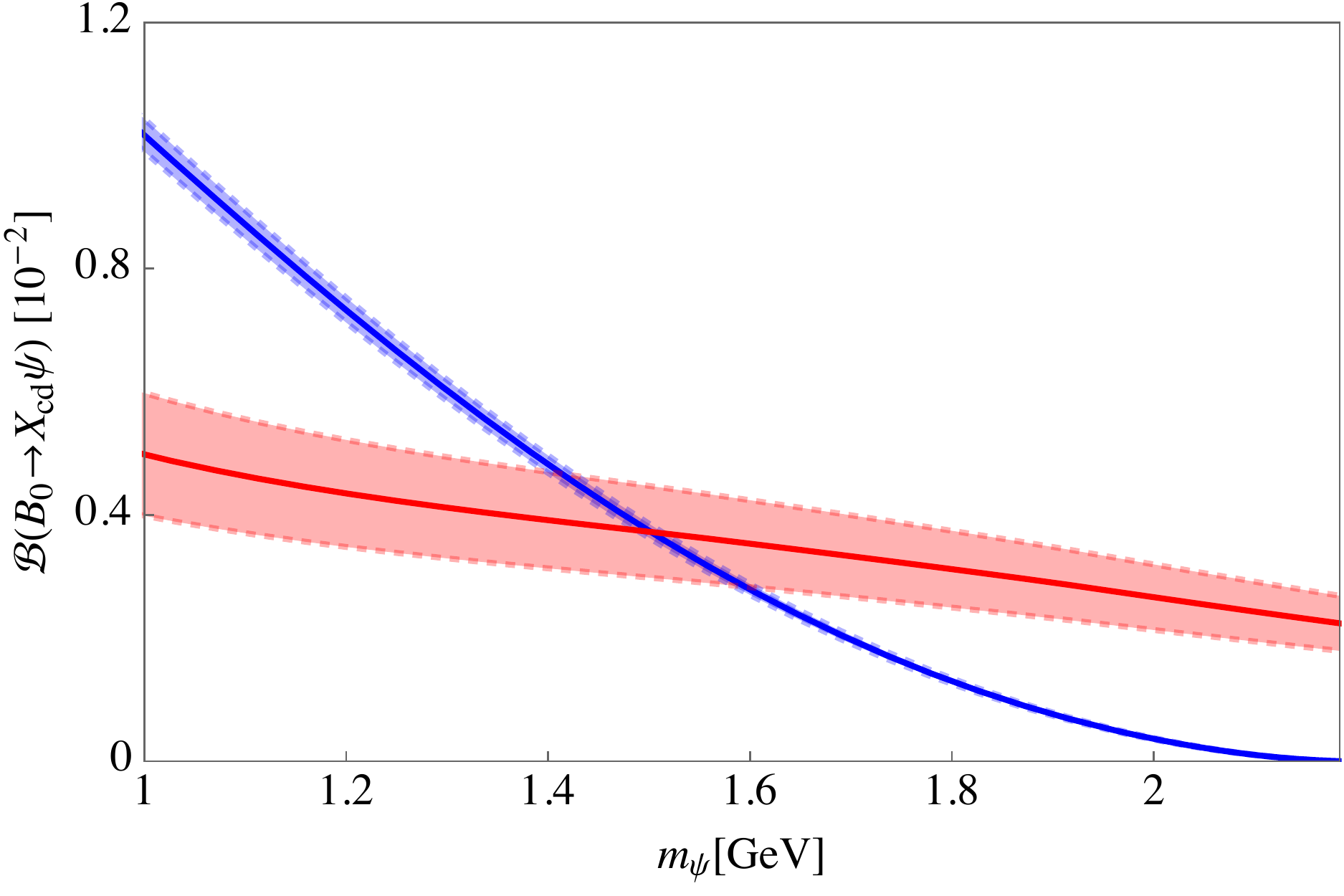} 
\includegraphics[width=0.45\columnwidth]{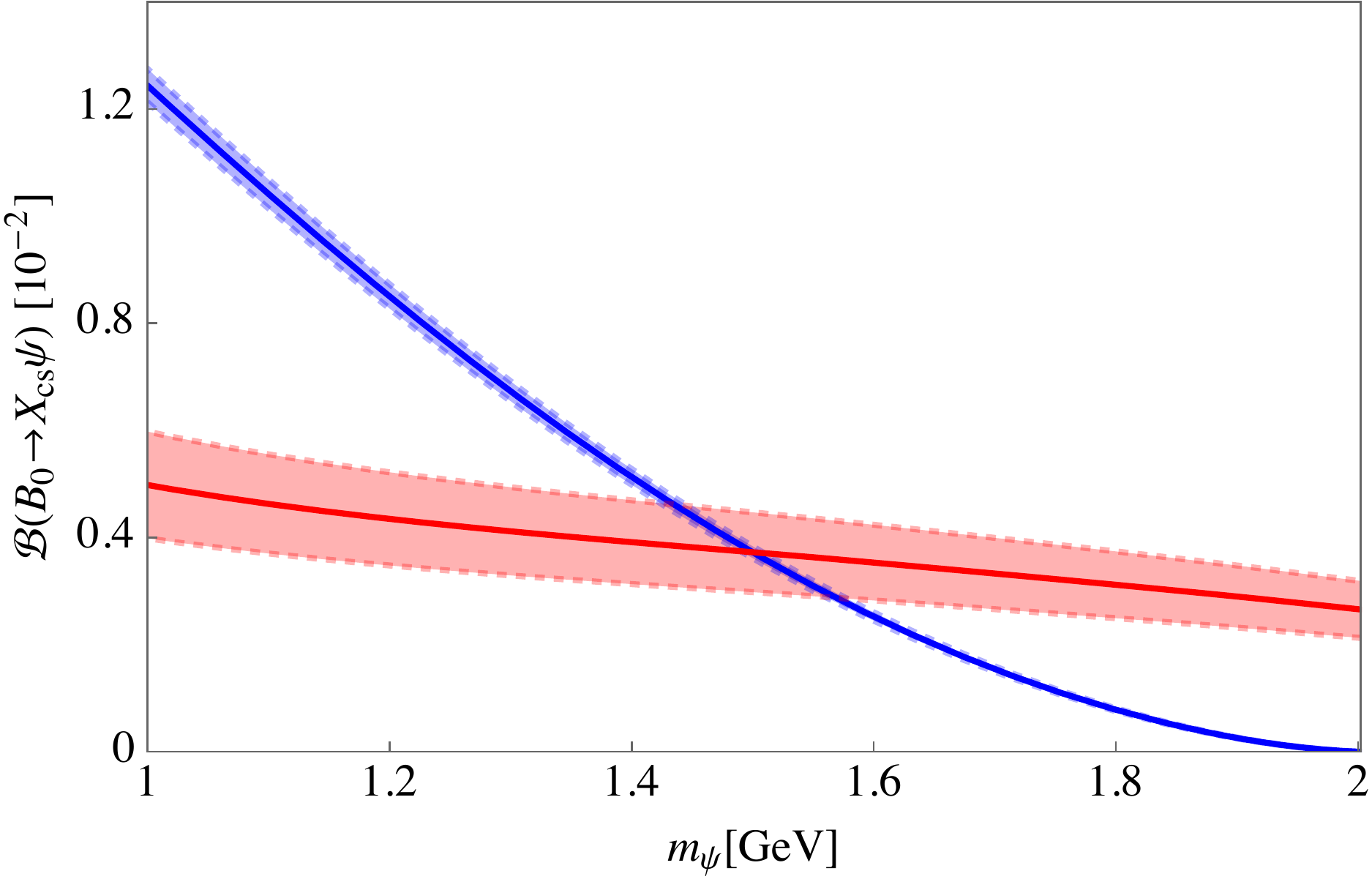} 
\caption{The red curves are 95$\%$ CL constraints on the inclusive $B$ meson decays into baryons and missing energy are estimated according to the ALEPH analysis given in the Ref.~\cite{Alonso-Alvarez:2021qfd}. The corresponding error bands come from the $20\%$ QCD corrections, which is a inferred percentage ratio according to the estimation on the exclusive decay branching fraction given in Eq.~(39) of Ref.\cite{Alonso-Alvarez:2021qfd}. The blue curves are the branching fractions calculated in the type-I model, where the $G_{uq}$ are taken as their upper limit values given in Eq.~(\ref{eq:Guqupper}): $G_{ud}^2=1.8\times 10^{-14} {\rm GeV}^{-4}$,  $G_{us}^2=3.75\times 10^{-14} {\rm GeV}^{-4}$, $G_{cd}^2=1.06\times 10^{-12} {\rm GeV}^{-4}$ and $G_{cs}^2=1.63\times 10^{-12} {\rm GeV}^{-4}$.}
\label{fig:B0toXuqpsiUpperTypeI} 
\end{center}
\end{figure}
\begin{figure}
\begin{center}
\includegraphics[width=0.45\columnwidth]{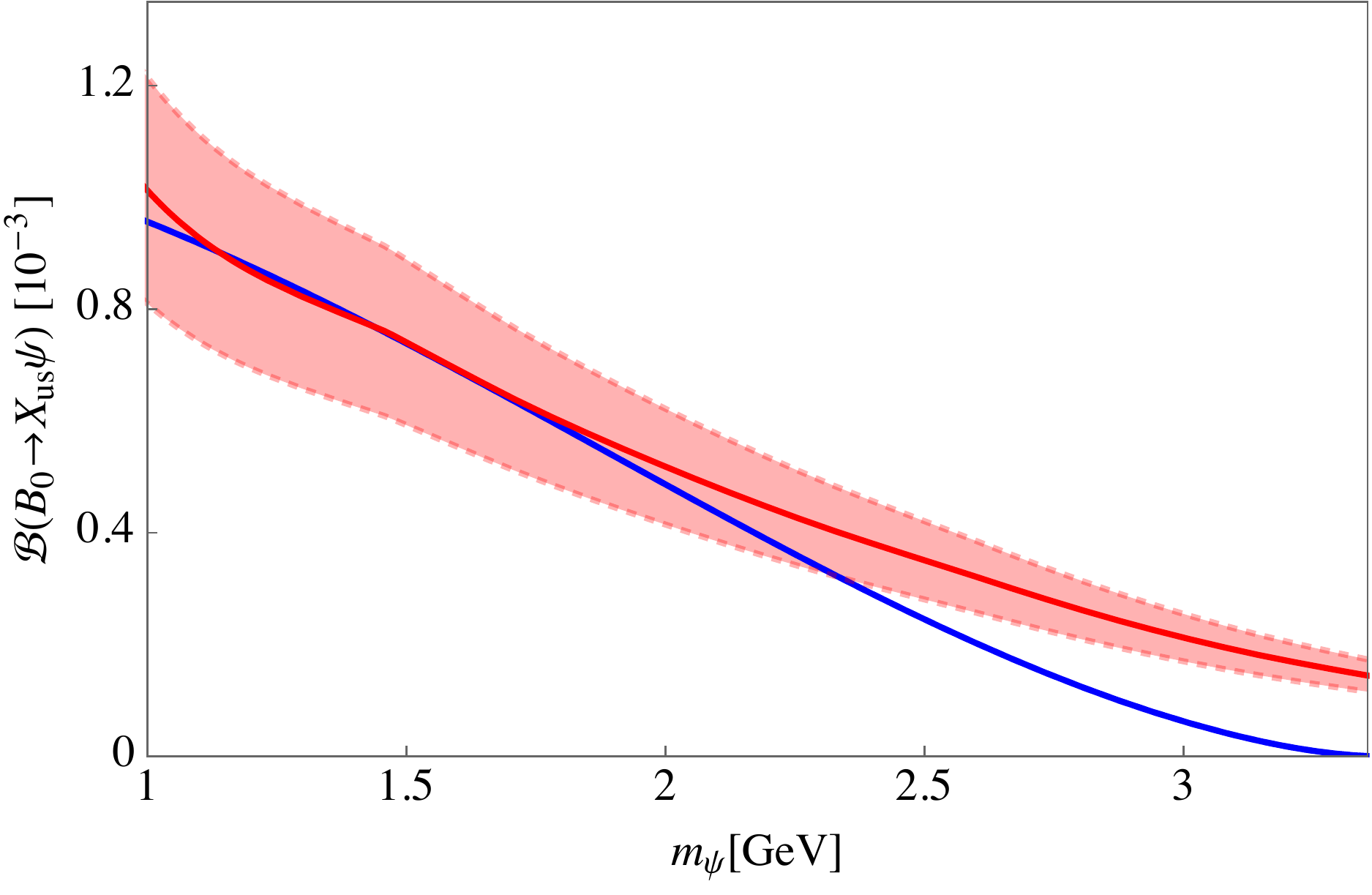}  
\includegraphics[width=0.45\columnwidth]{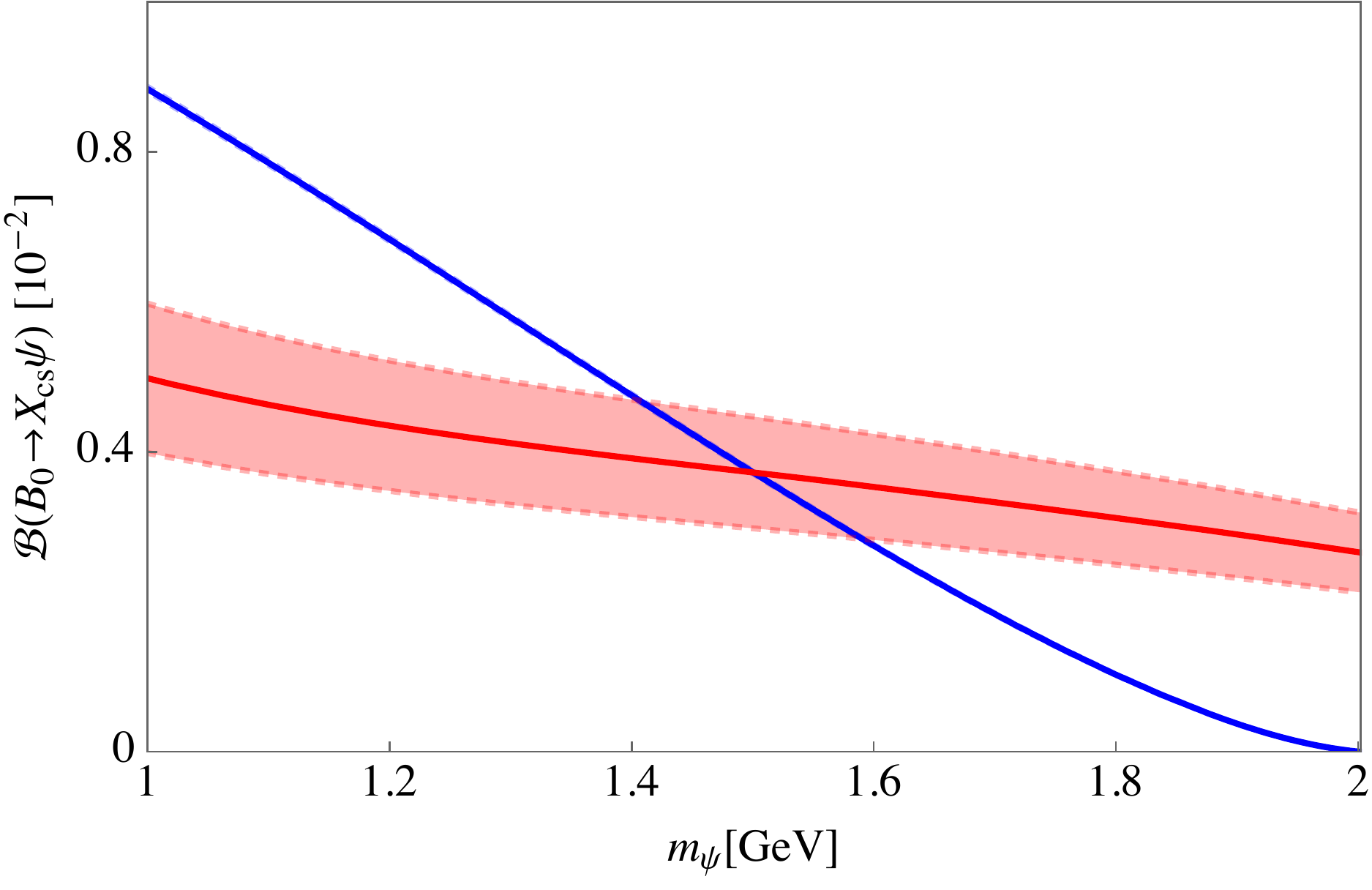} 
\caption{The red curves are the same as those in Fig.~\ref{fig:B0toXuqpsiUpperTypeI}. The blue curves are the branching fractions calculated in the type-II model, where the $G_{uq}$ are taken as their upper limit values given in Eq.~(\ref{eq:Guqupper}): $G_{us}^2=1.07\times 10^{-11} {\rm GeV}^{-4}$ and $G_{cs}^2=3.62\times 10^{-10} {\rm GeV}^{-4}$.}
\label{fig:B0toXuqpsiUpperTypeII} 
\end{center}
\end{figure}
In the Ref.~\cite{Alonso-Alvarez:2021qfd},  95$\%$ CL constraints on the inclusive $B$ meson decays into baryons and missing energy are estimated according to the ALEPH analysis \cite{ALEPH:2000vvi}, which is shown by the red curves in Fig.~\ref{fig:B0toXuqpsiUpperTypeI} and Fig.~\ref{fig:B0toXuqpsiUpperTypeII}. The corresponding error bands come from the $20\%$ QCD corrections, which is a inferred percentage ratio according to the estimation on the exclusive decay branching fraction given in the Eq.(39) of Ref.~\cite{Alonso-Alvarez:2021qfd}. On the other hand, it can be found that the branching fractions decrease with the increasing of $m_{\psi}$ as shown in Fig.~\ref{fig:B0toXuqpsiTypeI} and Fig.~\ref{fig:B0toXuqpsiTypeII}. Therefore, using the minimal possible value of $m_{\psi}$, one can in principle obtain the upper limits of the coupling constants $G_{uq}$. The restriction on $m_{\psi}$ given by \cite{Elor:2018twp} reads as: $1.5 {\rm GeV}<m_{\psi}<4.2 {\rm GeV}$. Setting $m_{\psi}$ at its minimum value: $m_{\psi}=1.5$ GeV for the red bands in Fig.~\ref{fig:B0toXuqpsiUpperTypeI}, we can obtain the upper limits and the corresponding errors for the branching fractions as: 
\begin{align} 
{\cal B}(B\to X_{ud}\psi)&<(3.73\pm0.75)\times 10^{-4},\nonumber\\
{\cal B}(B\to X_{us}\psi)&<(7.4\pm1.5)\times 10^{-4},\nonumber\\
{\cal B}(B\to X_{cd}/X_{cs}\psi)&<(3.72\pm0.74)\times 10^{-3}.\label{eq:constraints}
\end{align}
At the point $m_{\psi}=1.5$ GeV, comparing the center values of the branching fractions given in Fig.~\ref{fig:B0toXuqpsiTypeI} and Fig.~\ref{fig:B0toXuqpsiTypeII} with the constraints given above we can obtain the upper limits of $G_{uq}$ as
\begin{align} 
{\rm Type \ I}:~~~~& G_{ud}^2<(1.8\pm0.35)\times 10^{-14} {\rm GeV}^{-4},~~~G_{us}^2<(3.75\pm0.74)\times 10^{-14} {\rm GeV}^{-4},\nonumber\\
& G_{cd}^2<(1.06\pm 0.21)\times 10^{-12} {\rm GeV}^{-4},~~~G_{cs}^2<(1.63\pm0.33)\times 10^{-12} {\rm GeV}^{-4};\nonumber\\
{\rm Type\ II}:~~~~& G_{us}^2<(1.07\pm 0.21) \times 10^{-11} {\rm GeV}^{-4},~~~ G_{cs}^2<(3.62\pm0.72)\times 10^{-10} {\rm GeV}^{-4}.\label{eq:Guqupper}
\end{align}
Using the maximum value of $G_{uq}^2$ given above, we also present the branching fraction curves in Fig.~\ref{fig:B0toXuqpsiUpperTypeI} and Fig.~\ref{fig:B0toXuqpsiUpperTypeII}.  It can be found that with this setting of $G_{uq}$ the branching fraction curves are safely below the upper limit curves in the region $m_{\psi}>1.5$ GeV. Since the upper limit curves (red band) have much larger error than those of the branching fraction curves (blue band). To obtain the error for the constraints of $G_{uq}$ in Eq.~(\ref{eq:Guqupper}), we just compare the center value of the branching fraction with the upper and lower bounds of the red band at $m_{\psi}=1.5$ GeV.

Note that the branching fractions of $B_0\to X_{ud}\psi$ and $B_0\to X_{cd}\psi$ vanish in the type-II model due to the chiral limit, thus they cannot be used to constrain $G_{ud}^2$ and $G_{cd}^2$. The branching fractions of $B_0\to X_{us}\psi$ and $B_0\to X_{cs}\psi$ are suppressed by $m_s^2$, so they produce larger upper limits for $G_{uq}^2$.

\section{Conclusion}
\label{sec:conclusion}
In this work, using the recently developed $B$-Mesogenesis scenario, we have studied the semi-inclusive decays of $B$ meson into a dark anti-baryon $\psi$ plus any possible states $X$ containing $u/c$ and $d/s$ quarks with unit baryon number. The two types of effective Lagrangians proposed by the scenario are both considered in this work. The semi-inclusive decay branching fractions of $B\to X \psi$  are calculated by the method of heavy quark expansion,  where the non-perturbative contributions from the matrix elements of dimension-5 operators are included. We obtained the branching fractions as functions of the dark anti-baryon mass. Using the experimental  upper limits of the branching fractions, we provided the upper limits on the coupling constants in the  $B$-Mesogenesis scenario. In the Type-I model, the upper limits on $G_{ud}^2$ and $G_{us}^2$ are around $10^{-14}{\rm GeV}^{-4}$, while the upper limits on $G_{cd}^2$ and $G_{cs}^2$ are around $10^{-12}{\rm GeV}^{-4}$. The upper limits on $G_{us}^2$ and $G_{cs}^2$ in the Type-II model are around $10^{-11}{\rm GeV}^{-4}$ and $10^{-10}{\rm GeV}^{-4}$ respectively.

\section*{Acknowledgements}
The work of Y.J. Shi is supported by Opening Foundation of Shanghai Key Laboratory of Particle Physics and Cosmology under Grant No.22DZ2229013-2. The work of Y. Xing is supported by National Science Foundation of China under Grant No.12005294.  The work of Z.P. Xing is supported by China Postdoctoral Science Foundation under Grant No.2022M72210.

\begin{appendix}
	
\section{Two-body phase space integration}\label{2bdInt}

In this work, the rank-0 two-body phase space integration is defined as
\begin{align}
&\int \frac{d^4 l_1}{(2\pi)^3}\frac{d^4 l_2}{(2\pi)^3}\delta^4(P-l_1-l_2)\delta(l_1^2-m_1^2)\delta(l_2^2-m_2^2)\nonumber\\
=& \frac{\pi\sqrt{(P^2-(m_1+m_2)^2)(P^2-(m_1-m_2)^2)}}{2(2\pi)^6 P^2}.
\end{align}
the rank-2 two-body phase space integration is defined as
\begin{align}
&\int \frac{d^4 l_1}{(2\pi)^3}\frac{d^4 l_2}{(2\pi)^3}\delta^4(P-l_1-l_2)\delta(l_1^2-m_1^2)\delta(l_2^2-m_2^2)l_1^{\mu}l_2^{\nu}\nonumber\\
=& A_{\rm 2bd}[P^2,m_1^2,m_2^2]P^{\mu}P^{\nu}+B_{\rm 2bd}[P^2,m_1^2,m_2^2] P^2 g^{\mu\nu},
\end{align}
with 
\begin{align}
&A_{\rm 2bd}=\frac{\pi\sqrt{(P^2-(m_1+m_2)^2)(P^2-(m_1-m_2)^2)}}{6(2\pi)^6 (P^2)^3}\left[m_1^4+m_1^2(P^2-2m_2^2)+(P^2-2m_2^2)^2\right],\nonumber\\
&B_{\rm 2bd}=-\frac{\pi\left[(P^2-(m_1+m_2)^2)(P^2-(m_1-m_2)^2)\right]^{3/2}}{24(2\pi)^6 (P^2)^3}.
\end{align}

\section{Expressions of $A_{1,2}$ in the type-I model}\label{A1A2expr}
For convenience we define the following dimensionless variables to simplify the expressions: $\epsilon=E/m_b,\  s=m_q/m_b,\  u=m_u/m_b,\  q_s=Q^2/m_b^2$ and $\Delta=q_s^2-2 q_s (s+u)+(s-u)^2$. The expressions of $A_{1,2}$ in the type-I model read as
\begin{align}
A_1^{k^0}=&\frac{m_B m_b \sqrt{\Delta}}{48 \pi ^2 q_s^3}(\epsilon-1)\left(q_s^2+q_s (s-2
   u)+(s-u)^2\right),\\
A_2^{k^0}=&-\frac{m_b^2\sqrt{\Delta}}{192 \pi ^2 q_s^3}\left[q_s^2 (4 \epsilon-2 (s+u+2))+q_s \left(s (4
   \epsilon-2 u-4)+u (-8 \epsilon+u+8)+s^2\right)\right.\nonumber\\
   &\left.+4 (\epsilon-1)
   (s-u)^2+q_s^3\right];\\
A_1^{k^1}=&\frac{m_B}{96 \pi ^2 q_s^4\sqrt{\Delta}}A\left[2 q_s^3 u \left(3 \epsilon^2-6
   \epsilon-s+3 u+3\right)-q_s^2 \left(-s u \left(6 \epsilon^2-12
   \epsilon+7 u+6\right)\right.\right.\nonumber\\
   &\left.\left.+2 u^2 \left(9 \epsilon^2-18 \epsilon+2
   u+9\right)+s^3+2 s^2 u\right)+q_s (s-u)^2 \left(2 s \left(3
   \epsilon^2-6 \epsilon-u+3\right)\right.\right.\nonumber\\
   &\left.\left.+u \left(18 \epsilon^2-36
   \epsilon+u+18\right)+s^2\right)-6 (\epsilon-1)^2
   (s-u)^4+q_s^5-q_s^4 (s+4 u)\right]\nonumber\\
   &+\frac{m_B}{384 \pi ^2 m_b q_s^4\sqrt{\Delta}}(Y-Z)\left[-12 \epsilon^2 q_s^3 u-12
   \epsilon^2 q_s^2 s u+36 \epsilon^2 q_s^2 u^2-12
   \epsilon^2 q_s s^3\right.\nonumber\\
   &\left.-12 \epsilon^2 q_s s^2 u+60 \epsilon^2
   q_s s u^2-36 \epsilon^2 q_s u^3+12 \epsilon^2 s^4-48
   \epsilon^2 s^3 u+72 \epsilon^2 s^2 u^2-48 \epsilon^2 s u^3\right.\nonumber\\
   &\left.+12
   \epsilon^2 u^4+24 \epsilon q_s^3 u+24 \epsilon q_s^2 s
   u-72 \epsilon q_s^2 u^2+24 \epsilon q_s s^3+24 \epsilon
   q_s s^2 u-120 \epsilon q_s s u^2\right.\nonumber\\
   &\left.+72 \epsilon q_s
   u^3-24 \epsilon s^4+96 \epsilon s^3 u-144 \epsilon s^2 u^2+96
   \epsilon s u^3-24 \epsilon u^4+7 q_s^5-7 q_s^4 s\right.\nonumber\\
   &\left.-19
   q_s^4 u+3 q_s^3 s^2-2 q_s^3 s u+15 q_s^3 u^2-12
   q_s^3 u-q_s^2 s^3+q_s^2 s^2 u+q_s^2 s u^2-12
   q_s^2 s u\right.\nonumber\\
   &\left.-q_s^2 u^3+36 q_s^2 u^2-2 q_s s^4+8
   q_s s^3 u-12 q_s s^3-12 q_s s^2 u^2-12 q_s s^2
   u+8 q_s s u^3\right.\nonumber\\
   &\left.+60 q_s s u^2-2 q_s u^4-36 q_s
   u^3+12 s^4-48 s^3 u+72 s^2 u^2-48 s u^3+12 u^4\right],\\
A_2^{k^1}=&\frac{m_b}{192 \pi ^2 q_s^4\sqrt{\Delta}}A \left[q_s^3 \left(3 u \left(-4
   \epsilon^2+\epsilon (5 u+8)-9 u-4\right)+3 (\epsilon-1) s^2-2
   (\epsilon-3) s u\right)\right.\nonumber\\
   &\left.+q_s^2 \left(s u \left(-12 \epsilon^2+19
   \epsilon u+24 \epsilon-33 u-12\right)+u^2 \left(36
   \epsilon^2-\epsilon (13 u+72)+21 u+36\right)\right.\right.\nonumber\\
   &\left.\left.+(9-7 \epsilon)
   s^3+(\epsilon+3) s^2 u\right)+2 q_s (s-u)^2 \left(-2 s \left(3
   \epsilon^2+2 \epsilon (u-3)-3 u+3\right)\right.\right.\nonumber\\
   &\left.\left.+u \left(-18 \epsilon^2+2
   \epsilon (u+18)-3 (u+6)\right)+(2 \epsilon-3)
   s^2\right)\right.\nonumber\\
   &\left.+(\epsilon-3) q_s^5+q_s^4 ((15-7 \epsilon)
   u-(\epsilon-3) s)+12 (\epsilon-1)^2 (s-u)^4\right]\nonumber\\
   &+\frac{1}{384 \pi ^2 q_s^4\sqrt{\Delta}}(Y-Z) \left[12 \epsilon^2 q_s^3 u+12 \epsilon^2
   q_s^2 s u-36 \epsilon^2 q_s^2 u^2+12 \epsilon^2 q_s
   s^3+12 \epsilon^2 q_s s^2 u\right.\nonumber\\
   &\left.-60 \epsilon^2 q_s s u^2+36
   \epsilon^2 q_s u^3-12 \epsilon^2 s^4+48 \epsilon^2 s^3 u-72
   \epsilon^2 s^2 u^2+48 \epsilon^2 s u^3-12 \epsilon^2 u^4\right.\nonumber\\
   &\left.-\epsilon
   q_s^5+\epsilon q_s^4 s+7 \epsilon q_s^4 u-3
   \epsilon q_s^3 s^2+2 \epsilon q_s^3 s u-15 \epsilon
   q_s^3 u^2-24 \epsilon q_s^3 u+7 \epsilon q_s^2
   s^3-\epsilon q_s^2 s^2 u\right.\nonumber\\
   &\left.-19 \epsilon q_s^2 s u^2-24
   \epsilon q_s^2 s u+13 \epsilon q_s^2 u^3+72 \epsilon
   q_s^2 u^2-4 \epsilon q_s s^4+16 \epsilon q_s s^3
   u-24 \epsilon q_s s^3\right.\nonumber\\
   &\left.-24 \epsilon q_s s^2 u^2-24
   \epsilon q_s s^2 u+16 \epsilon q_s s u^3+120 \epsilon
   q_s s u^2-4 \epsilon q_s u^4-72 \epsilon q_s u^3+24
   \epsilon s^4\right.\nonumber\\
   &\left.-96 \epsilon s^3 u+144 \epsilon s^2 u^2-96 \epsilon s
   u^3+24 \epsilon u^4-6 q_s^5+6 q_s^4 s+12 q_s^4 u+12
   q_s^3 u-6 q_s^2 s^3\right.\nonumber\\
   &\left.+18 q_s^2 s u^2+12 q_s^2 s
   u-12 q_s^2 u^3-36 q_s^2 u^2+6 q_s s^4-24 q_s
   s^3 u+12 q_s s^3+36 q_s s^2 u^2\right.\nonumber\\
   &\left.+12 q_s s^2 u-24
   q_s s u^3-60 q_s s u^2+6 q_s u^4+36 q_s u^3-12
   s^4+48 s^3 u-72 s^2 u^2\right.\nonumber\\
   &\left.+48 s u^3-12 u^4\right];\\
A_1^{k^2}=&\frac{m_B}{32 \pi ^2 m_b q_s^5\Delta^{3/2}}Y(1-\epsilon)\left[q_s^5 u \left(4
   \epsilon^2-8 \epsilon-5 s-u+4\right)\right.\nonumber\\
   &\left.-3 q_s^4 \left(2 u^2
   \left(4 \epsilon^2-8 \epsilon+u+4\right)+s^3-3 s
   u^2\right)+q_s^3 \left(s^3 \left(8 \epsilon^2-16 \epsilon-11
   u+8\right)\right.\right.\nonumber\\
   &\left.\left.-3 s u^2 \left(8 \epsilon^2-16 \epsilon+7 u+8\right)+14 u^3
   \left(4 \epsilon^2-8 \epsilon+u+4\right)+9 s^4+9 s^2
   u^2\right)\right.\nonumber\\
   &\left.-q_s^2 (s-u)^2 \left(s^2 \left(24 \epsilon^2-48
   \epsilon-7 u+24\right)+s u \left(32 \epsilon^2-64 \epsilon-13
   u+32\right)\right.\right.\nonumber\\
   &\left.\left.+u^2 \left(64 \epsilon^2-128 \epsilon+11 u+64\right)+9
   s^3\right)+3 q_s (s-u)^4 \left(2 s \left(4 \epsilon^2-8
   \epsilon-u+4\right)\right.\right.\nonumber\\
   &\left.\left.+u \left(12 \epsilon^2-24
   \epsilon+u+12\right)+s^2\right)-8 (\epsilon-1)^2 (s-u)^6+q_s^6
   u\right],\\
A_2^{k^2}=&\frac{1}{128 \pi ^2 q_s^5\Delta^{3/2}}Y\left[4 q_s^6 \left(u \left(2 \epsilon^2-3
   \epsilon+u+1\right)+s^2\right)\right.\nonumber\\
   &\left.-12 (\epsilon-1) q_s (s-u)^4
   \left(-2 s \left(4 \epsilon^2+\epsilon (u-8)-2 u+4\right)\right.\right.\nonumber\\
   &\left.\left.+u \left(-12
   \epsilon^2+\epsilon (u+24)-2 (u+6)\right)+(\epsilon-2) s^2\right)-2
   q_s^5 \left(s^2 \left(2 \epsilon^2-4 \epsilon-3 u+2\right)\right.\right.\nonumber\\
   &\left.\left.+s u
   \left(4 \epsilon^2+2 \epsilon-3 (u+2)\right)+u \left(-8
   \epsilon^3+\epsilon^2 (22 u+24)-6 \epsilon (7 u+4)+3 u^2+20
   u+8\right)+3 s^3\right)\right.\nonumber\\
   &\left.+3 q_s^4 \left(4 s^3 \left(2 \epsilon^2-5
   \epsilon-u+3\right)-4 s u^2 \left(2 \epsilon^2-7\epsilon+u+5\right)\right.\right.\nonumber\\
   &\left.\left.+u^2 \left(-32 \epsilon^3+32 \epsilon^2 (u+3)-24
   \epsilon (3 u+4)+3 u^2+40 u+32\right)+3 s^4+2 s^2
   u^2\right)\right.\nonumber\\
   &\left.-q_s^3 \left(3 s^4 \left(16 \epsilon^2-44 \epsilon-7
   u+28\right)+2 s^2 u^2 \left(12 \epsilon^2-42 \epsilon+7 u+30\right)\right.\right.\nonumber\\
   &\left.\left.-2
   s^3 \left(16 \epsilon^3+16 \epsilon^2 (u-3)-6 \epsilon (9 u-8)-7
   u^2+38 u-16\right)\right.\right.\nonumber\\
   &\left.\left.-3 s u^2 \left(-32 \epsilon^3+48 \epsilon^2 (u+2)-4
   \epsilon (31 u+24)+7 u^2+76 u+32\right)\right.\right.\nonumber\\
   &\left.\left.+u^3 \left(-224 \epsilon^3+8
   \epsilon^2 (13 u+84)-24 \epsilon (11 u+28)+7 u^2+160 u+224\right)+7
   s^5\right)\right.\nonumber\\
   &\left.+2 q_s^2 (s-u)^2 \left(s^3 \left(20 \epsilon^2-58
   \epsilon-4 u+38\right)\right.\right.\nonumber\\
   &\left.\left.-2 s^2 \left(24 \epsilon^3+6 \epsilon^2
   (u-12)+\epsilon (72-19 u)-3 u^2+13 u-24\right)\right.\right.\nonumber\\
   &\left.\left.-2 s u \left(32
   \epsilon^3+6 \epsilon^2 (3 u-16)+\epsilon (96-49 u)+2 u^2+31
   u-32\right)\right.\right.\nonumber\\
   &\left.\left.+u^2 \left(-128 \epsilon^3+4 \epsilon^2 (7 u+96)-6
   \epsilon (13 u+64)+u^2+50 u+128\right)+s^4\right)\right.\nonumber\\
   &\left.-32 (\epsilon-1)^3
   (s-u)^6+q_s^8-3 q_s^7 (s+u)\right];\\
A_1^{ug}=&-\frac{3m_B}{64\pi ^2 m_b q_s^3\sqrt{\Delta}}N(1-\epsilon)(q_s+s-u)\left[q_s^2-q_s (s+2 u)+(s-u)^2\right],\\
A_2^{ug}=&-\frac{3}{256 \pi ^2 q_s^3\sqrt{\Delta}}N(q_s+s-u)\left[4 \epsilon q_s^2-4 \epsilon q_s s-8 \epsilon q_s u+4
   \epsilon s^2-8 \epsilon s u+4 \epsilon u^2+q_s^3\right.\nonumber\\
   &\left.-2 q_s^2
   s-2 q_s^2 u-4 q_s^2+q_s s^2-2 q_s s u+4
   q_s s+q_s u^2+8 q_s u-4 s^2+8 s u-4 u^2\right];\\
A_1^{qg}=&\frac{m_B}{32\pi ^2 m_b q_s^3\sqrt{\Delta}}N(1-\epsilon)(q_s-s+u)\left[q_s^2-q_s (2s+u)+(s-u)^2\right],\\
A_2^{qg}=&\frac{1}{128 \pi ^2 q_s^3\sqrt{\Delta}}N(q_s-s+u)\left[4 \epsilon q_s^2-8 \epsilon q_s s-4 \epsilon q_s u+4
   \epsilon s^2-8 \epsilon s u+4 \epsilon u^2+q_s^3\right.\nonumber\\
   &\left.-2 q_s^2
   s+2 q_s^2 u-4 q_s^2+q_s s^2-2 q_s s u+8
   q_s s+q_s u^2+4 q_s u-4 s^2+8 s u-4 u^2\right].
\end{align}
Here the superscript $ug$ and $qg$ denote the cases of gluon emission from the $u$ and $q$ quarks respectively.

\section{Expressions of $A_{1,2}$ in the type-II model}\label{A1A2exprTypeII}
For convenience we define the following dimensionless variables to simplify the expressions: $\epsilon=E/m_b,\  s=m_q/m_b,\  u=m_u/m_b,\  q_s=Q^2/m_b^2$ and $\Delta=q_s^2-2 q_s (s+u)+(s-u)^2$. The expressions of $A_{1,2}$ in the type-II model read as
\begin{align}
A_1^{k^0}=&0,\\
A_2^{k^0}=&\frac{m_b^2 s \sqrt{\Delta}}{16\pi ^2 q_s};\\
A_1^{k^1}=&\frac{m_B}{32\pi ^2 m_b q_s^2\sqrt{\Delta}}(Y-Z)\left[q_s(s+u)-(s-u)^2\right],\\
A_2^{k^1}=&\frac{1}{32\pi ^2 q_s^2\sqrt{\Delta}}\left[2(\epsilon-1)m_b A+\epsilon(Z-Y)\right]\left[q_s(s+u)-(s-u)^2\right];\\
A_1^{k^2}=&0,\\
A_2^{k^2}=&\frac{s}{32\pi ^2 q_s^3{\Delta}^{3/2}} Y\left[-4 \epsilon^2 q_s^3 s-4 \epsilon^2 q_s^3 u+12 \epsilon^2
   q_s^2 s^2+12 \epsilon^2 q_s^2 u^2-12 \epsilon^2 q_s
   s^3+12 \epsilon^2 q_s s^2 u\right.\nonumber\\
   &\left.+12 \epsilon^2 q_s s u^2-12
   \epsilon^2 q_s u^3+4 \epsilon^2 s^4-16 \epsilon^2 s^3 u+24
   \epsilon^2 s^2 u^2-16 \epsilon^2 s u^3+4 \epsilon^2 u^4\right.\nonumber\\
   &\left.+8 \epsilon
   q_s^3 s+8 \epsilon q_s^3 u-24 \epsilon q_s^2 s^2-24
   \epsilon q_s^2 u^2+24 \epsilon q_s s^3-24 \epsilon
   q_s s^2 u-24 \epsilon q_s s u^2+24 \epsilon q_s
   u^3\right.\nonumber\\
   &\left.-8 \epsilon s^4+32 \epsilon s^3 u-48 \epsilon s^2 u^2+32
   \epsilon s u^3-8 \epsilon u^4+q_s^4 s+q_s^4 u-3
   q_s^3 s^2+6 q_s^3 s u-4 q_s^3 s\right.\nonumber\\
   &\left.-3 q_s^3 u^2-4
   q_s^3 u+3 q_s^2 s^3-3 q_s^2 s^2 u+12 q_s^2
   s^2-3 q_s^2 s u^2+3 q_s^2 u^3+12 q_s^2 u^2-q_s
   s^4\right.\nonumber\\
   &\left.+4 q_s s^3 u-12 q_s s^3-6 q_s s^2 u^2+12 q_s
   s^2 u+4 q_s s u^3+12 q_s s u^2-q_s u^4-12 q_s
   u^3+4 s^4\right.\nonumber\\
   &\left.-16 s^3 u+24 s^2 u^2-16 s u^3+4 u^4\right],
\end{align}
and $A_{1,2}^{ug}=A_{1,2}^{qg}=0$ in the chiral limit. 

\end{appendix}

\end{document}